\documentclass[a4paper,11pt]{article}
\usepackage{jheppub}
\usepackage{subcaption}

\preprint{}
\affiliation{$^{1}$Department of physics, Shanghai University, Shanghai, 200444, China}
\affiliation{$^{2}$School of Physics, University of Chinese Academy of Sciences, Beijing, 100049, China}
\affiliation{$^{3}$Center for Gravitation and Cosmology, Yangzhou University, Yangzhou 225009, China}
\emailAdd{guzhuofan@shu.edu.cn}
\emailAdd{yanyukun20@mails.ucas.ac.cn}
\emailAdd{sfwu@shu.edu.cn}

\begin{document}

\title{\boldmath
Neural ODEs for holographic transport models without translation symmetry }
\author{Zhuo-Fan Gu$^{1}$, Yu-Kun Yan$^{2}$, and Shao-Feng Wu$^{1,3}$}

\abstract{
We investigate the data-driven holographic transport models without
translation symmetry, focusing on the real part of frequency-dependent shear viscosity, $\eta_{\mathrm{re}}(\omega)$. We develop a
radial flow equation of the shear response and establish its relation to $\eta _{\mathrm{re}}(\omega)$ for a wide class of holographic models. This allows
us to determine $\eta _{\mathrm{re}}(\omega )$ of a strongly coupled field
theory by the black hole metric and the graviton mass. The latter serves as
the bulk dual to the translation symmetry breaking on the boundary. We
convert the flow equation to a Neural Ordinary Differential Equation (Neural
ODE), which is a neural network with continuous depth and produces output
through a black-box ODE solver. Testing the Neural ODE on three well-known holographic models without translation symmetry, we demonstrate its ability to accurately learn either the metric or mass when given the other. Additionally, we illustrate that the learned metric can be used to predict the
derivative of entanglement entropy $S$ with respect to the size of entangling region $l$.
}

\flushbottom

\maketitle

\pagebreak

\section{Introduction}

The AdS/CFT correspondence is an elegant holographic mapping that brings
illuminating insights into quantum gravity and provides fruitful phenomenal
models of strongly coupled quantum many-body systems \cite%
{Aharony9905,Liu2004}. The standard approach to the holographic modeling
involves positing a bulk action, solving the Einstein equation to derive the
bulk metric, and then using the holographic dictionary to transform various
gravitational observables into field-theory observables. However,
constructing a suitable action from a phenomenological perspective is a
fundamental and challenging inverse problem in theoretical physics. In
addition to relying on the symmetry principle, it usually requires deep
domain knowledge and intuition from human experts.

In certain conditions within the realm of holography, some field-theory
observables can be derived directly from the bulk metric without
necessitating the complete action. This leads to a reduced inverse problem:
Can we reconstruct the bulk metric from these field-theory observables? This
question is central to an important branch of the broader bulk
reconstruction project \cite{Harlow:2018fse,DeJonckheere:2017qkk,Kajuri2003}%
, which ambitiously aims to reorganize the degrees of freedom in the CFT
into local gravitational physics under specific limits. There are many
different methods for reconstructing the bulk metric, part of which can be
found in \cite%
{deHaro:2000vlm,Hammersley:2006cp,Hubeny:2006yu,Hammersley:2007ab,Hubeny:2012ry,Bilson:2008ab,Bilson:2010ff,Balasubramanian:2013lsa,Myers:2014jia,Czech:2014ppa,Engelhardt:2016wgb,Engelhardt:2016crc,Roy:2018ehv,Kabat:2018smf,Hashimoto:2020mrx,Caron-Huot2211,Hashimoto:2021umd,Jokela:2020auu,Jokela2304,XWB2023,Qi2306,Yang2023}%
. Among others, Hashimoto et al. propose an intriguing scheme using the
neural network and deep learning (DL) \cite{Hashimoto:2018ftp}. They
discretize the Klein-Gordon equation for a scalar field in curved spacetime
into the architecture of a neural network. The data they use comprises the
expectation value of the operator and its conjugate source, with the IR
behavior of the field serving as labels. Under the inductive bias that the
metric should be smooth, they design a regularization term to guide the
training process. After training, the bulk metric is encoded as the weights
of the neural network. This scheme is termed the AdS/DL correspondence. It
has been quickly applied to construct holographic QCD models and produced
interesting outcomes, such as extracting novel features from lattice data on
chiral condensate, predicting reasonable masses for excited-state mesons,
and deriving the explicit form of a dilaton potential \cite%
{Hashimoto:2018bnb,Akutagawa:2020yeo,Hashimoto:2021ihd}.

In addition to QCD, holographic models have found extensive applications in
condensed matter physics, hydrodynamics, and quantum information. Notably,
transport phenomena, such as momentum and charge transports, are considered
very suitable for the holographic research based on the bottom-up setup. In
this context, Yan et al. propose a data-driven holographic transport model 
\cite{Wu2020}. Their data are the complex frequency-dependent shear
viscosity and the network architecture corresponds to the discretization of
a radial flow equation. Since the IR solution of the flow equation is fully
determined by regularity, the forward propagation of information is designed
from IR to UV. Compared to previous algorithms, this design eliminates the
need for additional IR labels, reducing systematic errors in the analysis.
Recently, a similar algorithm has been developed to learn from optical
conductivity \cite{Li:2022zjc}, which has the potential applications in
condensed matter physics, such as strange metal and high-temperature
superconductivity.

All the aforementioned works on AdS/DL share a common feature: the emergent
extra dimension corresponds to the depth of the neural network. In fact, the
neural network itself is regarded as a bulk spacetime. However, since the
neural network has discrete layers, both the spacetime metric and the
equation of motion are discretized. This leads to intrinsic discretization
errors in the algorithm. Additionally, the regularization term used to
induce smooth metrics increases both arbitrariness and computational cost.

To address these discretization problems, Neural ODEs have been introduced
for AdS/DL \cite{Hashimoto:2020,Hashimoto:2022eij,Ahn2401}. This is a relatively new
family of deep neural network models that generalize the standard
layer-to-layer propagation to continuous-depth models \cite{Chen:2018}. In
essence, Neural ODEs employ a neural network to parameterize the derivative
of the hidden state, while the output is computed using a black-box adaptive
differential equation solver. The application of Neural ODEs in holographic
modeling, as highlighted in \cite{Hashimoto:2020}, showcases the potential
of AI as more than just a data processing tool in theoretical physics, but
also as a valuable aid in scientific discoveries.

The purpose of this paper is twofold. On a technical level, we aim to
replace the discrete architecture in \cite{Wu2020} with Neural ODEs. Note
that the metric ansatz was limited to a polynomial in \cite%
{Hashimoto:2020,Hashimoto:2022eij,Ahn2401}, which constrains the
representation capability of Neural ODEs. In this work, we will eliminate
this constraint by incorporating the neural network into the ansatz. It
should be pointed out that switching from polynomials to neural networks
increases the training difficulty significantly. Fortunately, we find an
effective training method. The key strategies include using multiple
optimizers in sequence, as well as employing a larger number of initial
values and the early-stopping technique to prevent Neural ODEs from getting
stuck with bad parameters. On a physical level, we intend to break the
translation symmetry that is respected by the theoretical framework in \cite%
{Wu2020}. The latter is indispensable to describe the real-world condensed
matter. However, the breaking of translation symmetry presents an important
challenge. Following the spirit of AdS/DL, one would expect the framework of
the theory to be as universal as possible, so that the machine has the
potential to learn novel metrics from the data of field theories, where the
gravity dual may not be explicit. In \cite{Hashimoto:2018ftp}, the
universality is achieved under the probe limit, where the bulk action
generating the background metric can be entirely unknown. In \cite{Wu2020},
the probe limit is not invoked, but the universality depends heavily on the
relationship between the shear viscosity of the boundary field theory and
the shear response on the UV boundary. If the translation symmetry is
broken, this relationship usually no longer holds. In this work, through a
careful analysis of the UV behavior related to the holographic
renormalization \cite{Skenderis0209,Papadimitriou2016}, we find that the
shear viscosity and the shear response on the boundary usually have the same
real part, even if the translation symmetry is broken. This crucial
observation motivates us to use only the real part of the shear viscosity as
the training data\footnote{%
It is instructive to note that the shear viscosity does not have the hydrodynamic interpretation when the translation symmetry is broken,  
but it still quantifies the entropy production rate due to the strain \cite{Hartnoll1601}. In fact, the definition of
the viscosity tensor is independent with the hydrodynamics: it is the linear
response of the stress tensor under a time--dependent deformation \cite%
{Read1207}. Recent studies on the viscosity for the systems without
translation symmetry can be found, e.g., in \cite{Burmistrov1901,Baggioli1910}.%
}.

Once the data is prepared, we will define our learning tasks. Following \cite%
{Hartnoll1601}, we will focus on the holographic models that break
translation symmetry but maintain homogeneity and isotropy in geometry. In
many of these models, we will demonstrate that the real part of the
frequency-dependent shear viscosity is determined exclusively by functions
of the bulk metric and the graviton mass. As a reverse engineering, our
objective is to investigate whether Neural ODEs can learn one of these
functions when the other is given\footnote{%
Since there is a trade-off between metric and mass in determining shear
viscosity, we do not attempt to learn the two functions simultaneously.}.

The rest of this paper is organized as follows. In Section 2, we will
develop a radial flow equation for a large class of the holographic models
without translation symmetry. In Section 3, we will introduce the machine
learning algorithm, including the architecture, loss function and inductive
bias. In Section 4, we will generate the data and display the training
results based on three well-known holographic models without translation
symmetry. The conclusion and discussion will be presented in Section 5.
Additionally, there are four appendices. In Appendix A, we will establish
the relationship between the shear viscosity and the shear response. In
Appendix B, we will provide the training scheme and report. In Appendix C,
we will improve the performance of Neural ODEs at low temperatures. In
Appendix D, we will study the entanglement entropy (EE) on the boundary
using the bulk metric learned by Neural ODEs.

\section{Holographic flow equation}

Consider a strongly coupled field theory that is holographically dual to the
3+1 dimensional Einstein gravity minimally coupled with the matter. Suppose
that it allows a planar black hole solution, which is described by the line
element%
\begin{equation}
ds^{2}=-g_{tt}(r)dt^{2}+g_{rr}(r)dr^{2}+g_{xx}(r)d\vec{x}^{2},
\label{BH ansatz}
\end{equation}%
and sourced by the energy-momentum tensor%
\begin{equation}
T_{\mu \nu }=\mathrm{diag}\left(
T_{tt}(r),T_{rr}(r),T_{xx}(r),T_{xx}(r)\right) .  \label{Tuv}
\end{equation}%
Importantly, eq. (\ref{BH ansatz}) and eq. (\ref{Tuv}) are homogeneous and
isotropic along the field theory directions, although we have not assumed
that the matter fields are homogeneous. When the black hole is perturbed by
the time-dependent shear mode $\delta g_{xy}=g_{xx}h(r)e^{-i\omega t}$, the
wave equation has a general form \cite{Hartnoll1601}\footnote{%
Here we have assumed that this mode decouples from other perturbations.
Since $\delta g_{xy}$ transforms as a parity-even tensor mode under the
rotation in $x-y$ plane, this assumption is quite general. However, one can
find a counterexample in the Einstein-$SU(2)$ theory \cite{Son1311}, where
the coupling appears from a particular background that locks the symmetries
in spacetime and $SU(2)$ vector space.}%
\begin{equation}
\frac{1}{\sqrt{-g}}\partial _{r}(\sqrt{-g}g^{rr}\partial _{r}h)+\left(
\omega ^{2}g^{tt}-m^{2}\right) h=0,  \label{wave eq0}
\end{equation}%
where $m$ is the radially varying graviton mass 
\begin{equation}
m\left( r\right) ^{2}=g^{xx}T_{xx}-\frac{\delta T_{xy}}{\delta g_{xy}},
\label{m2}
\end{equation}
and the Newton constant has been set as 16$\pi G_{N}=1$. Note that the
nonzero graviton mass in the bulk is dual to the breaking of the translation
symmetry on the boundary \cite{ZaanenBook}.

Consider the bulk spacetime sliced along the radial direction and introduce
the shear response function at each slice%
\begin{equation}
\chi =\frac{\Pi }{i\omega h},  \label{chi0}
\end{equation}%
where $\Pi =-\sqrt{-g}g^{rr}\partial _{r}h$ is the momentum conjugate to the
field $h$ in the Hamiltonian formulation. Using eq. (\ref{chi0}), one can
rewrite the wave equation as%
\begin{equation}
\partial _{r}\chi -i\omega \sqrt{\frac{g_{rr}}{g_{tt}}}\left( \frac{\chi ^{2}%
}{g_{xx}}-g_{xx}\right) +\frac{1}{i\omega }\sqrt{g_{tt}g_{rr}}g_{xx}m^{2}=0,
\label{floweq1}
\end{equation}%
which is a radial flow equation of the shear response.

Observing the flow equation (\ref{floweq1}), one can see that the regularity
of the shear response indicates%
\begin{equation}
\chi (r_{\mathrm{h}})=g_{xx}(r_{\mathrm{h}}),  \label{chiIR}
\end{equation}%
where $r_{\mathrm{h}}$ is the horizon radius. Using eq. (\ref{chiIR}) as the
IR boundary condition, the flow equation can be solved to obtain the complex
shear response at UV. However, this shear response does not generally
represent the desired shear viscosity on the boundary field theory.
Nevertheless, through a careful UV analysis in Appendix A, we will
demonstrate that for a wide range of theories, their real parts are indeed
the same:%
\begin{equation}
\eta _{\mathrm{re}}\left( \omega \right) =\left. \chi _{\mathrm{re}}\left(
\omega \right) \right\vert _{r\rightarrow \infty }.  \label{UV relation}
\end{equation}%
We emphasize that this nontrivial observation is crucial to the theoretical
framework of our data-driven model.

To proceed, we will exploit the scaling symmetry of this system%
\begin{equation}
r\rightarrow r\lambda ,\ \left( t,x,y\right) \rightarrow \left( t,x,y\right)
/\lambda , \ g_{rr}\rightarrow g_{rr}\lambda ^{-2},\ g_{tt}\rightarrow
g_{tt}\lambda ^{2},\ g_{xx}\rightarrow g_{xx}\lambda ^{2},
\end{equation}%
which indicates that one can set $r_{\mathrm{h}}=1$ for convenience.

We will further simplify the metric (\ref{BH ansatz}) by imposing $%
g_{tt}g_{rr}=1$ and setting $g_{xx}=r^{2}$. Then the metric ansatz can be
written as%
\begin{equation}
ds^{2}=-r^{2}f(r)dt^{2}+\frac{1}{r^{2}f(r)}dr^{2}+r^{2}d\vec{x}^{2}.
\label{fmetric}
\end{equation}%
Replacing the radial coordinate with $z=1/r$, one can see that the horizon
is located at $z_{\mathrm{h}}=1$ and the boundary at $z_{\mathrm{b}}=0$.
Accordingly, the flow equation (\ref{floweq1}) can be reduced to\footnote{%
If $g_{tt}g_{rr}\neq 1$, this equation is still valid, but $f$ and $m^{2}$
will represent the joint factors $\frac{1}{r^{2}}\sqrt{\frac{g_{tt}(r)}{%
g_{rr}(r)}}$ and $\sqrt{g_{tt}(r)g_{rr}(r)}m(r)^{2}$, respectively.}%
\begin{equation}
\partial _{z}\chi +\frac{i\omega }{f(z)}\left( z^{2}\chi ^{2}-\frac{1}{z^{2}}%
\right) -\frac{1}{i\omega }\frac{m^{2}}{z^{4}}=0,  \label{floweq2}
\end{equation}%
and the IR boundary condition (\ref{chiIR}) and the UV relation (\ref{UV
relation}) are changed into $\chi (z_{\mathrm{h}})=1$ and $\eta _{\mathrm{re}%
}=\chi _{\mathrm{re}}\left( z_{\mathrm{b}}\right) $, respectively.

\section{Machine learning algorithm}

\subsection{Neural ODE}

In residual networks \cite{He1512}, there is a sequence of transformations
to a hidden state%
\begin{equation}
x_{t+1}=x_{t}+y_{t}(x_{t},\theta _{t}),
\end{equation}%
where $x_{t}$ is the hidden state at layer $t$, $y_{t}$\ is a differentiable
function preserves the dimension of $x_{t}$, and $\theta _{t}$ denotes the
transform parameters. The difference between $x_{t+1}$ and $x_{t}$ can be
interpreted as a discretization of the derivative $x^{\prime }(t)$ with the
step $\Delta t=1$. Adding more layers and decreasing the step, one can
approach the limit%
\begin{equation}
\frac{dx(t)}{dt}=y(x(t),t,\theta ).
\end{equation}%
This equation can be used to represent the Neural ODE, an ODE induced by the
continuous-depth limit of residual networks.

Given an initial state $x(t_{0})$, the forward propagation of Neural ODEs
yields the state $x(t_{1})$ through a black-box ODE solver. The back
propagation of Neural ODEs is realized by the adjoint sensitivity method 
\cite{Pontryagin:1962}. Suppose that there is a loss function depending on
the outputs of an ODE solver%
\begin{equation}
L\left( x\left( t_{1}\right) \right) =L(x\left( t_{0}\right)
+\int_{t_{0}}^{t_{1}}y(x(t),t,\theta )dt).
\end{equation}%
The adjoint state is defined as%
\begin{equation}
a(t)=\frac{\partial L}{\partial x\left( t\right) }.
\end{equation}%
It has been proven that the adjoint dynamics is dominated by another ODE%
\begin{equation}
\frac{da(t)}{dt}=-a(t)\cdot \frac{\partial y}{\partial x},
\end{equation}%
and the parameter gradient can be calculated by an integral \cite{Chen:2018}%
\begin{equation}
\frac{\partial L}{\partial \theta }=\int_{t_{1}}^{t_{0}}a(t)\cdot \frac{%
\partial y}{\partial \theta }dt.
\end{equation}

Now we can transform the flow equation into the Neural ODE. We first split
the flow equation (\ref{floweq2}) into real and imaginary parts%
\begin{eqnarray}
\frac{d\chi _{\mathrm{re}}}{dz} &=&\frac{2\omega z^{2}}{f}\chi _{\mathrm{re}%
}\chi _{\mathrm{im}}, \\
\frac{d\chi _{\mathrm{im}}}{dz} &=&\frac{\omega }{f}\left[ \left( z\chi _{%
\mathrm{im}}\right) ^{2}-\left( z\chi _{\mathrm{re}}\right) ^{2}+\frac{1}{%
z^{2}}\right] -\frac{m^{2}}{\omega z^{4}}.
\end{eqnarray}%
Then we identify the shear response $\left( \chi _{\mathrm{re}},\chi _{%
\mathrm{im}}\right) $ with the state $x$ and the radial coordinate $z$ with
the time $t$. In general, the function of metric $f(z)$ or the function of
mass square $m^{2}(z)$ can be represented by a neural network with the
parameter $\theta $.

\begin{figure}[h]
\centering
\includegraphics[width=13cm]{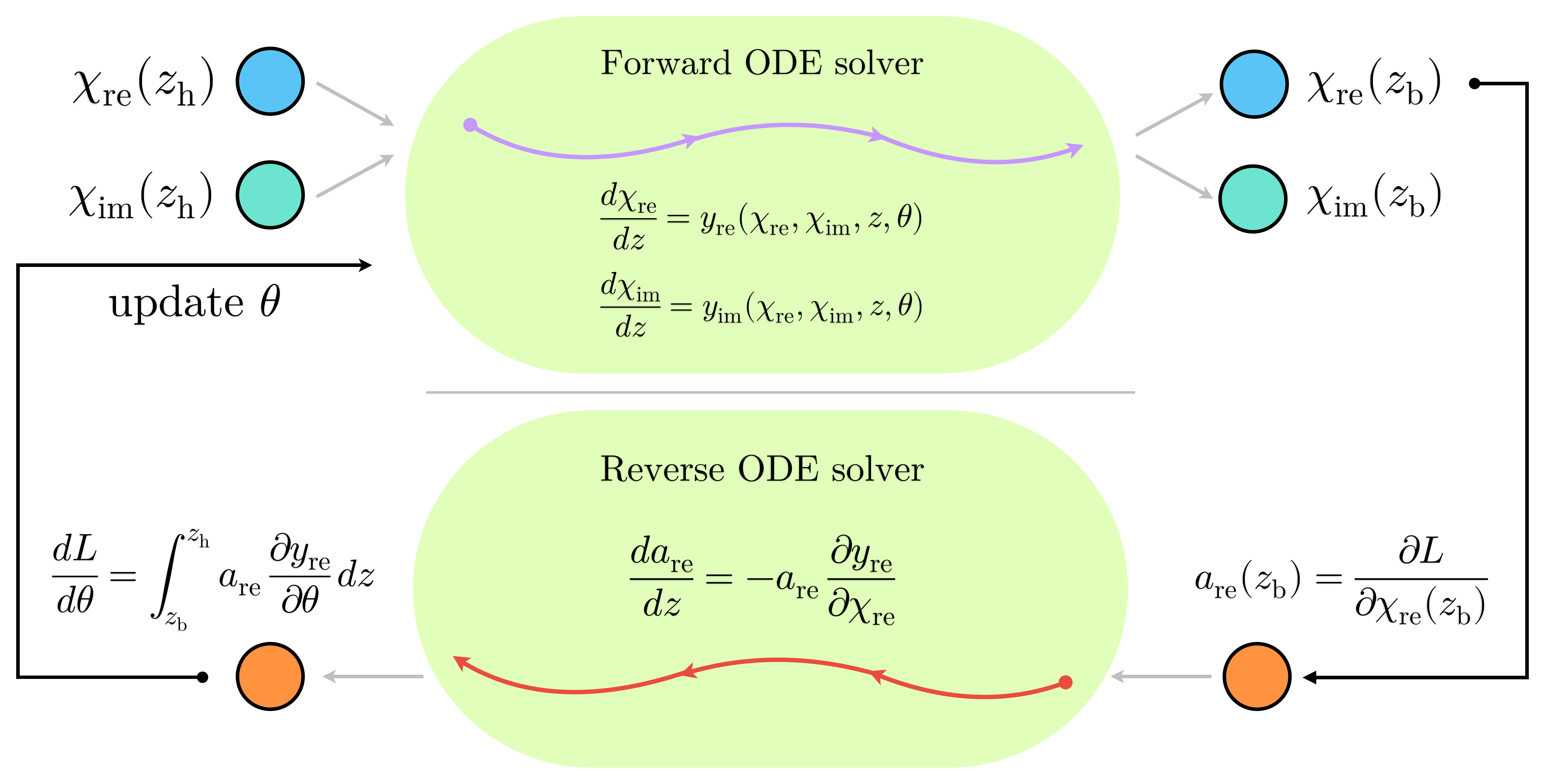}
\caption{The architecture of the algorithm. The upper part depicts the
forward propagation (from IR to UV) of the real and imaginary parts of the
shear response. The lower part depicts the backward propagation of the
adjoint state related only to the real part.}
\label{fig:NODE}
\end{figure}

Obviously, given the parameter $\theta $ is equivalent to giving the trial
function of metric or mass square which specifies the Neural ODE. By
invoking an ODE solver, one can integrate from IR to UV and predict the real
part of shear viscosity. Using the adjoint sensitivity method, one can
implement back-propagation and perform training. We depict the architecture
of the current algorithm in figure \ref{fig:NODE}. Note that it differs from
a typical Neural ODE in two aspects: the data is not a time series but only
relates to the output at the upper limit of the integration, and one
component of the adjoint state vanishes.

\subsection{Loss function and inductive bias}

Our loss function is chosen as the log-cosh form \cite%
{LC2018,LC2020,LC2021,LC2022}%
\begin{equation}
L=\frac{1}{N}\sum_{i=1}^{N}\log \left[ \cosh \left( \eta _{\mathrm{re}%
}\left( \omega _{i},\theta \right) -\bar{\eta}_{\mathrm{re}}\left( \omega
_{i}\right) \right) \right] ,
\end{equation}%
where $N$ denotes the size of data, $\bar{\eta}_{\mathrm{re}}$ represents
the input data, and $\eta _{\mathrm{re}}$ is what the Neural ODE predicts.
Note that the loss $\log \left[ \cosh \left( x\right) \right] $ behaves like
L1 loss $\left\vert x\right\vert $ far from the origin and L2 loss $x^{2}$
close to the origin. It is twice differentiable everywhere, making it
suitable for optimizers like BFGS that require the Hessian \cite{Nocedal}.

We need to impose some inductive biases, which depend on different tasks.

Task 1: Learning the metric

Suppose that the graviton mass is given. Since we focus on finite
temperature field theories, there should be a horizon at $z=1$. This bias is
implemented through the ansatz%
\begin{equation}
f(z)=(1-z)n_{1}(z,\theta _{1}),
\end{equation}%
where $n_{1}$ is a neural network with the trainable parameter $\theta _{1}$.

Task 2: Learning the mass

Suppose that the metric is given. We propose the ansatz%
\begin{equation}
m(z)^{2}=z^{b}n_{2}(z,\theta _{2})
\end{equation}%
where the exponent $b$ is a hyperparameter and $n_{2}$ is a neural network
with the trainable parameter $\theta _{2}$. The purpose of separating $z^{b}$
from $m(z)^{2}$ is to input the UV information $m(z)^{2}\sim z^{b}+\cdots $
through hyperparameter tuning, see Appendix B.

\section{Holographic models}

We will study three well-known holographic models without translation
symmetry, all of which have homogeneous and isotropic backgrounds.

\subsection{Massive gravity}

The research of massive gravity (MG) has a long and winding history \cite%
{Pauli1939,Veltman1970,Zakharov1970,Vainshtein1972,Boulware1972}. The main
theoretical interest is to explore whether there exists a self-consistent
theory that has a massive graviton with spin 2. In cosmology, MG is
considered as a candidate theory that explains the accelerated expansion of
the universe by modifying Einstein gravity. In holography, massive gravity
is the first analytically tractable model without translation symmetry \cite%
{Vegh1301}

We will consider the dRGT massive gravity that is claimed to be ghost free 
\cite{deRham2010,deRham2011,deRham1401}. Its simplest version used in
holography is described by the bulk action%
\begin{equation}
S_{\mathrm{bulk}}=\int d^{4}x\sqrt{-g}\left( \mathcal{R}+6-\alpha \mathrm{tr}%
\mathcal{X}\right) ,
\end{equation}%
where $\mathcal{X}{^{\mu }}_{\nu }=\sqrt{g^{\mu \lambda }f_{\lambda \nu }}$
and $f_{\mu \nu }=\mathrm{diag}(0,0,1,1)$ is the reference metric. Note that
we have set the AdS radius to one for convenience.

Using the action, one can derive the Einstein equation%
\begin{equation}
R_{\mu \nu }-\frac{1}{2}g_{\mu \nu }R-3g_{\mu \nu }=\frac{1}{2}\alpha \left( 
\mathcal{X}_{\mu \nu }-g_{\mu \nu }\mathrm{tr}\mathcal{X}\right) .
\end{equation}%
This allows for the existence of a black hole solution (\ref{fmetric}) with
the emblackening factor%
\begin{equation}
f(z)=1-z^{3}-\frac{\alpha }{2}z(1-z^{2}).  \label{metricMG}
\end{equation}%
The black hole is associated with the Hawking temperature%
\begin{equation}
T=\frac{1}{4\pi }\left( 3-\alpha \right) ,
\end{equation}%
which should be non-negative. Inserting the energy-momentum tensor%
\begin{equation}
T_{\mu \nu }=\alpha \left( \mathcal{X}_{\mu \nu }-g_{\mu \nu }\mathrm{tr}%
\mathcal{X}\right)
\end{equation}%
into eq. (\ref{m2}), one can calculate the square of graviton mass%
\begin{equation}
m^{2}=\frac{\alpha }{2}z.  \label{m2MG}
\end{equation}

\begin{figure}[h]
\centering
\begin{subfigure}{0.326\textwidth}
    	\includegraphics[width=\textwidth]{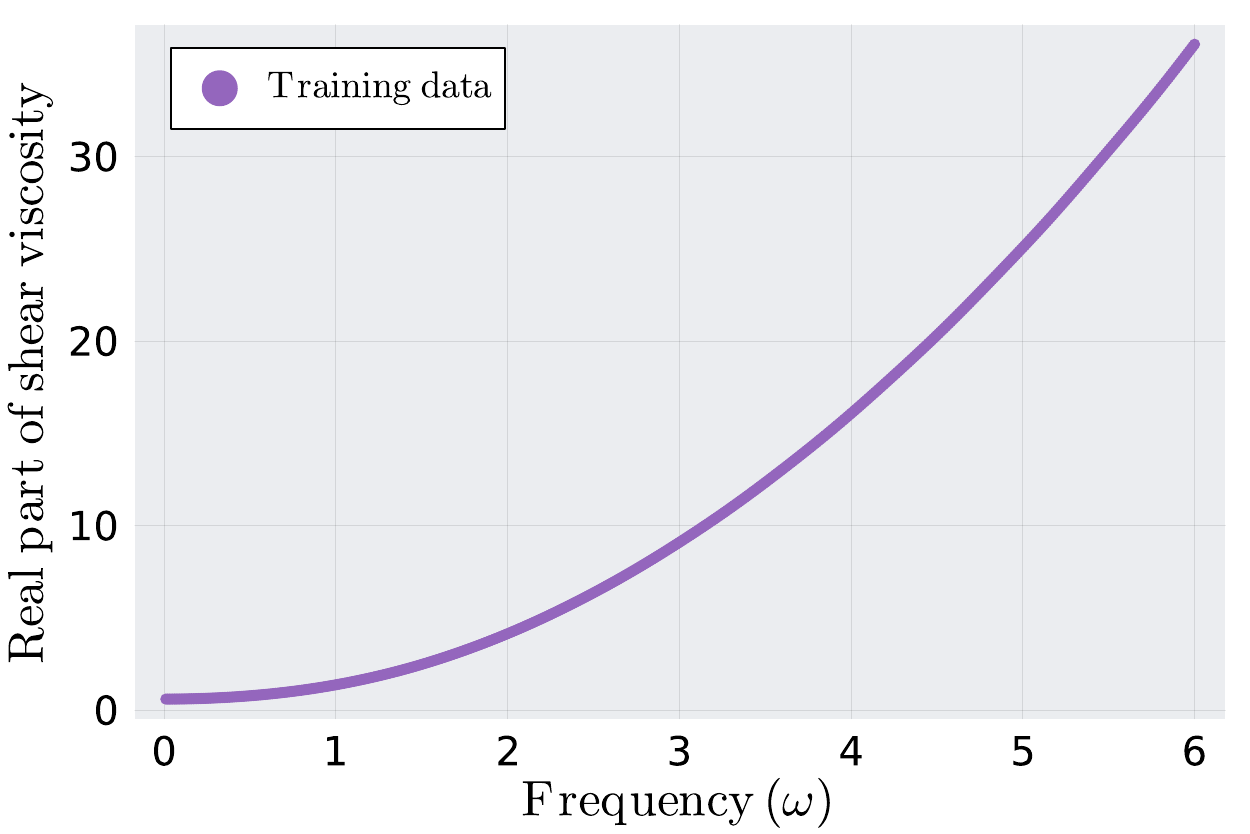}
  	\end{subfigure}
\begin{subfigure}{0.326\textwidth}
    	\includegraphics[width=\textwidth]{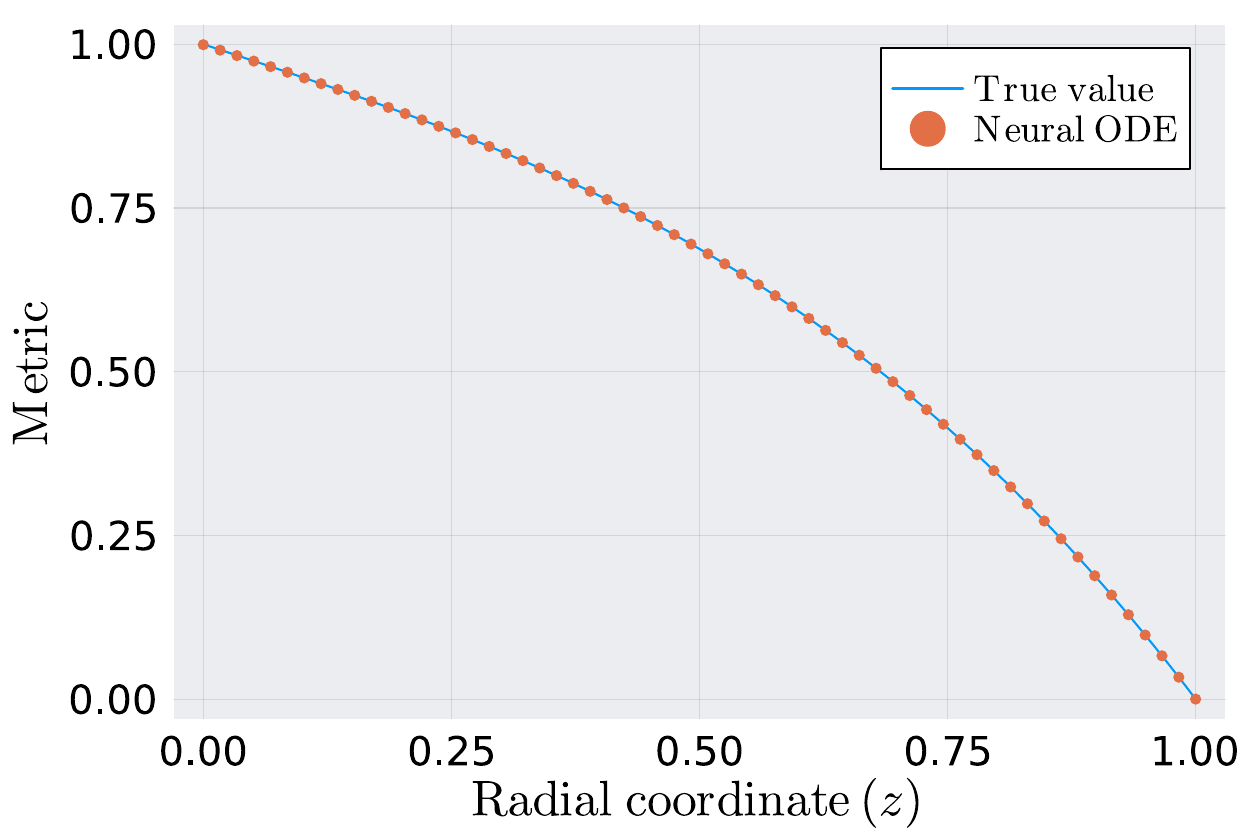}
  	\end{subfigure}
\begin{subfigure}{0.326\textwidth}
    	\includegraphics[width=\textwidth]{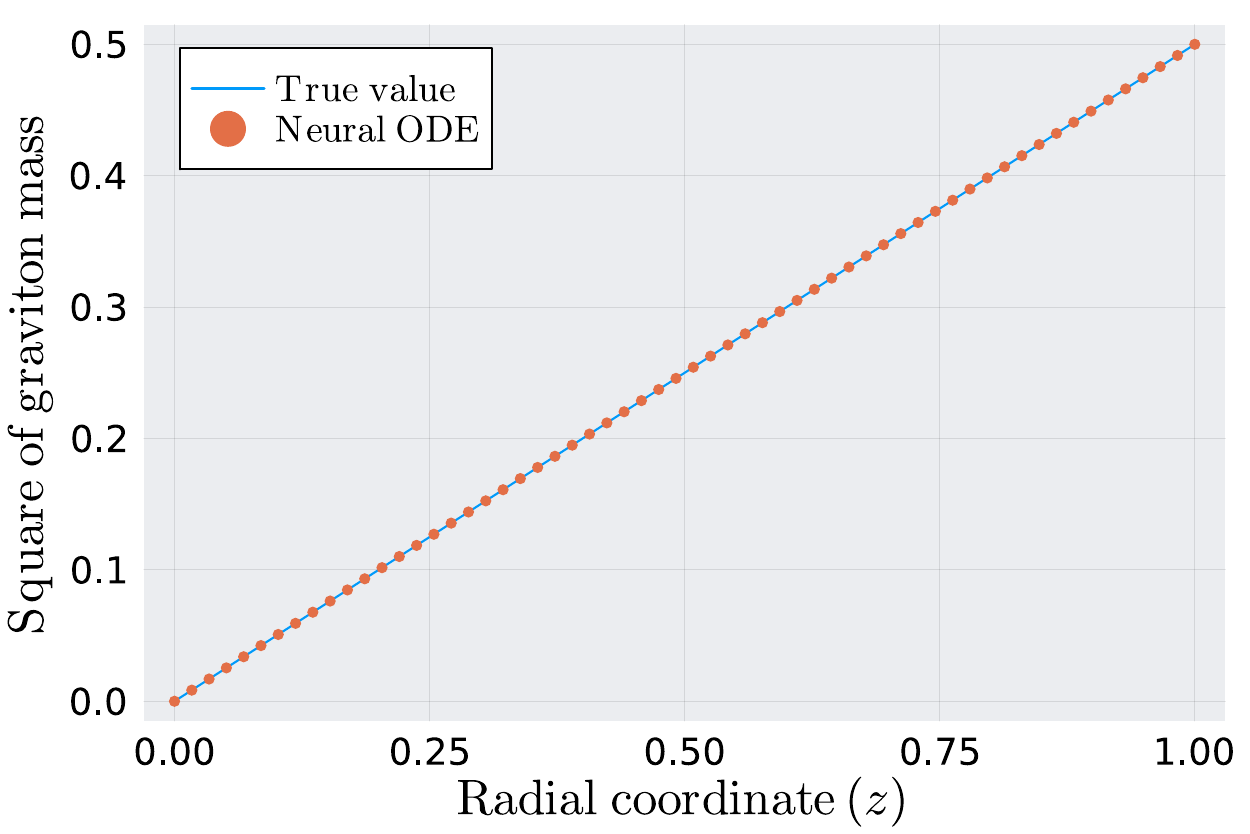}
  	\end{subfigure}
\caption{Training data and performance of Neural ODEs for the MG model. The
left panel shows the real part of the frequency-dependent shear viscosity,
obtained by solving the flow equation with the true metric and mass. The
middle and right panels illustrate the performance of Neural ODEs in
learning the metric and mass respectively, where the curves represent the
true values and the dots indicate the predictions.}
\label{fig:MG}
\end{figure}

Now we can generate the data for training. Inputting eq. (\ref{metricMG})
and eq. (\ref{m2MG}) into eq. (\ref{floweq2}), using the regular condition
at the horizon, and setting the parameter $\alpha =1$, one can solve the
flow equation and obtain the real part of shear viscosity on the boundary.
We sample the frequency uniformly between 0.01 and 6 to generate 600 data
points $\left[ \omega ,\eta _{\mathrm{re}}(\omega )\right] $, which are
plotted in the left panel of figure \ref{fig:MG}.

We conduct two numerical experiments: fixing the mass to learn the metric
and fixing the metric to learn the mass. The primary training results are
shown in the middle and right panels of figure \ref{fig:MG}. Additionally,
figure \ref{fig:LA_b} and table \ref{report1} in Appendix B detail the
hyperparameter tuning, minimum loss, and mean relative error (MRE).

\subsection{Linear axions}

The simplest holographic mechanism for breaking translation symmetry is to
invoke the linear axion (LA) \cite{Andrade:1311}. Consider the Einstein
gravity minimally coupled with two massless scalar fields. Its bulk action
is given by%
\begin{equation}
S_{\mathrm{bulk}}=\int d^{4}x\sqrt{-g}\left( \mathcal{R}+6-X\right) ,
\label{LA action}
\end{equation}%
where $X=g^{\mu \nu }X_{\mu \nu }$ and $X_{\mu \nu }=G_{IJ}\partial _{\mu
}\chi ^{I}\partial _{\nu }\chi ^{J}$ with $G_{IJ}=\frac{1}{2}\mathrm{diag}%
(1,1)$ and $I=1,2$.

The equations of motion following the action are%
\begin{eqnarray}
R_{\mu \nu }-\frac{1}{2}g_{\mu \nu }R-3g_{\mu \nu } &=&X_{\mu \nu }-\frac{1}{%
2}g_{\mu \nu }X, \\
\nabla ^{2}\chi ^{I} &=&0.
\end{eqnarray}%
They admit a black hole solution%
\begin{equation}
f(z)=1-z^{3}+\frac{\beta ^{2}}{2}z^{2}(z-1)  \label{LAmetric}
\end{equation}%
with the temperature%
\begin{equation}
T=\frac{1}{8\pi }(6-\beta ^{2}),
\end{equation}%
when the axions are linear%
\begin{equation}
\chi ^{1}=\beta x,\;\chi ^{2}=\beta y.
\end{equation}%
Using the energy-momentum tensor%
\begin{equation}
T_{\mu \nu }=2X_{\mu \nu }-g_{\mu \nu }X
\end{equation}%
and eq. (\ref{m2}), we read the square of graviton mass 
\begin{equation}
m^{2}=\beta ^{2}z^{2}.  \label{LAm2}
\end{equation}

\begin{figure}[h]
\centering
\begin{subfigure}{0.326\textwidth}
    	\includegraphics[width=\textwidth]{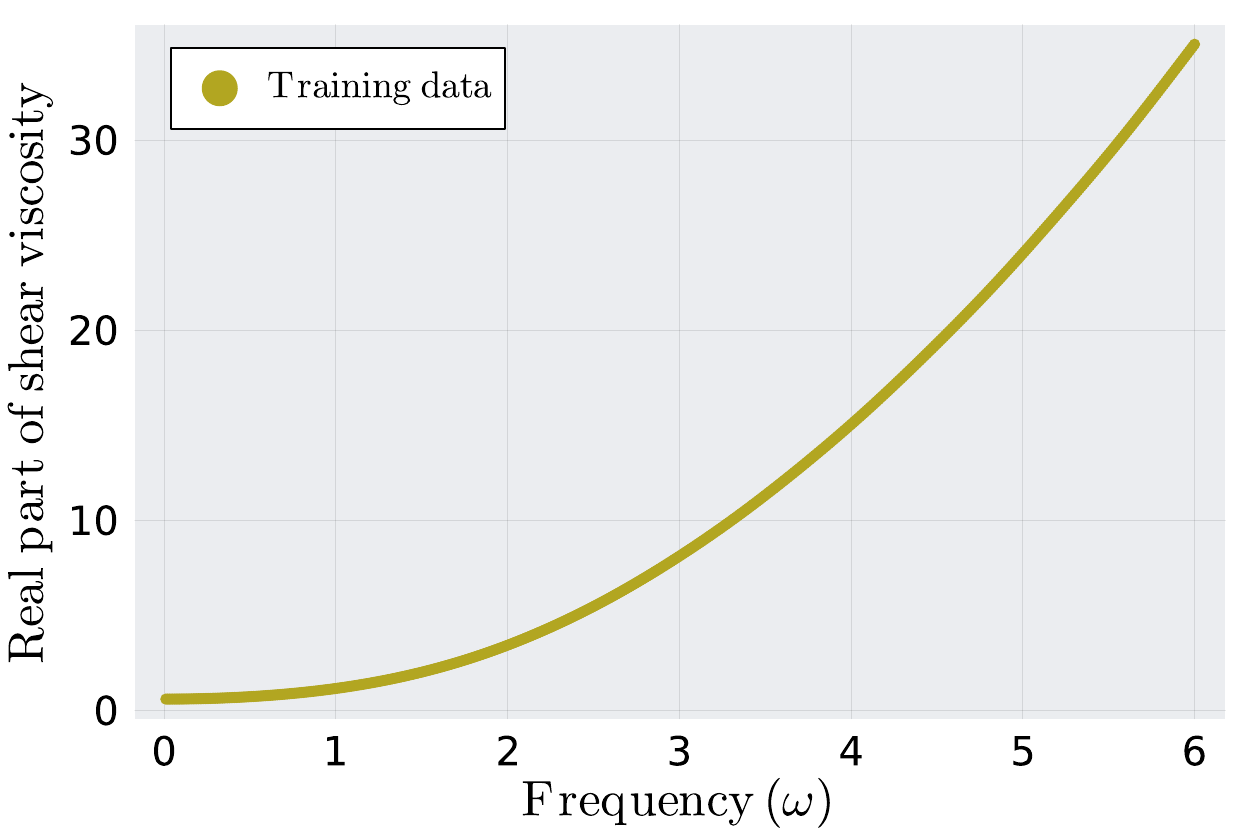}
  	\end{subfigure}
\begin{subfigure}{0.326\textwidth}
    	\includegraphics[width=\textwidth]{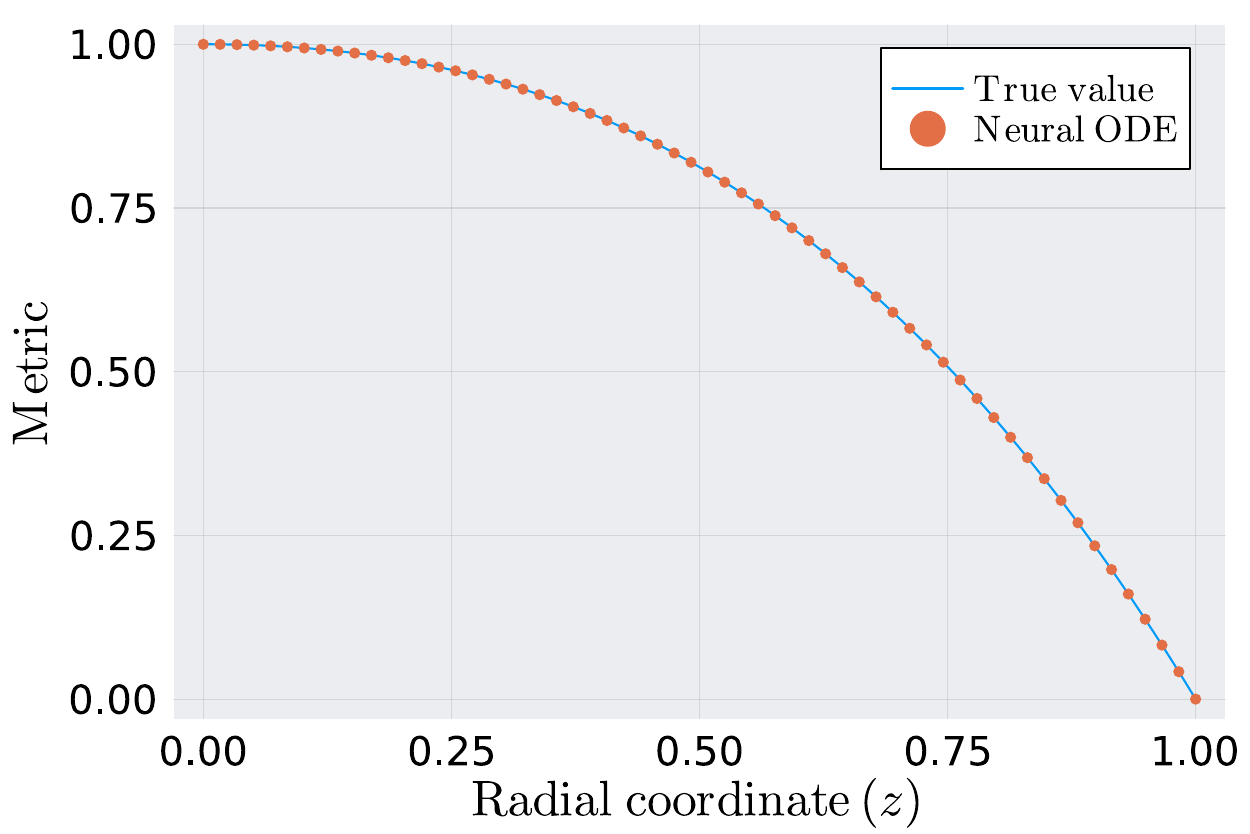}
  	\end{subfigure}
\begin{subfigure}{0.326\textwidth}
    	\includegraphics[width=\textwidth]{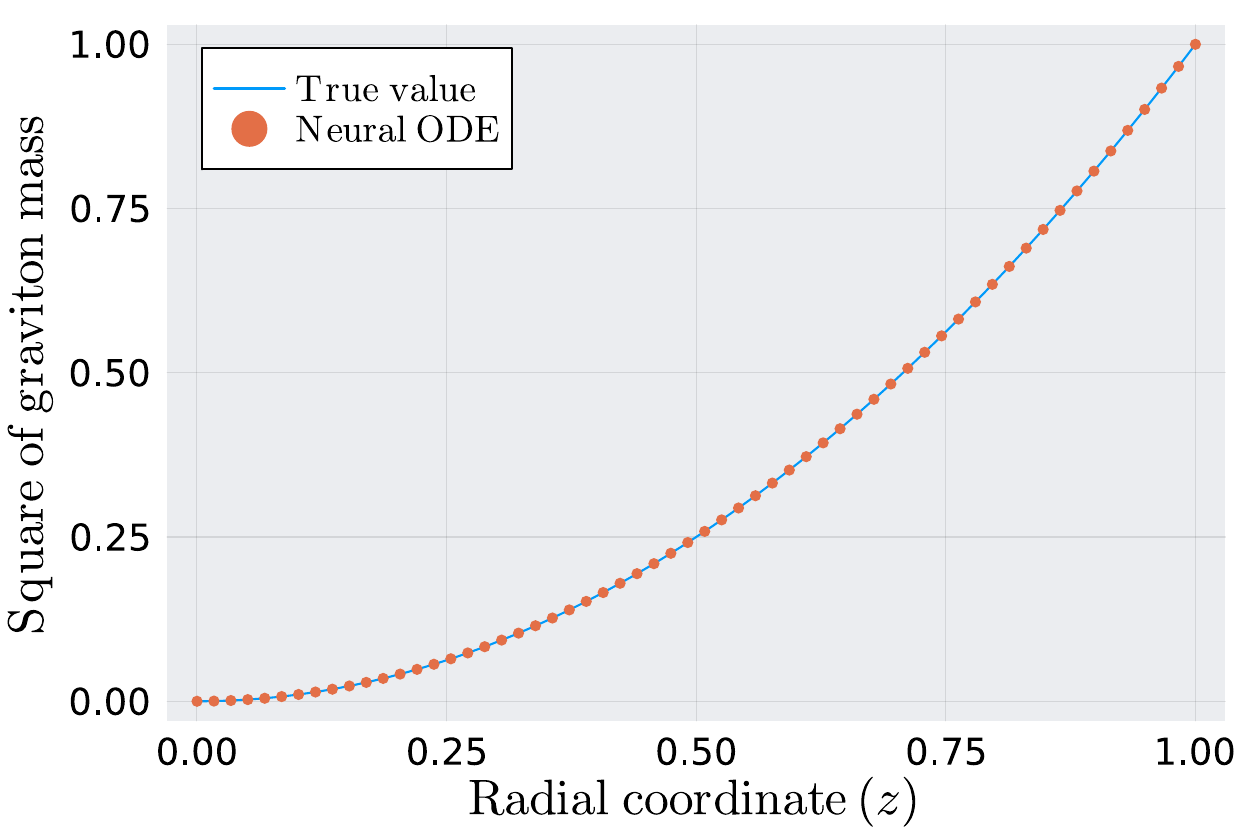}
  	\end{subfigure}
\caption{Training data and performance of Neural ODEs for the LA model. Left
panel: the real part of the frequency-dependent shear viscosity. Middle and
right panels: The performance of Neural ODEs in learning the metric and
mass. }
\label{fig:LA}
\end{figure}

With eq. (\ref{LAmetric}) and eq. (\ref{LAm2}) in hand, we can generate the
data. Here we fix the parameter $\beta =1$. Other setting is similar to
before. The data $\left[ \omega ,\eta _{\mathrm{re}}(\omega )\right] $ are
plotted in the left panel of figure \ref{fig:LA}. The training results are
displayed in the middle and right panels of figure \ref{fig:LA}, as well as
in figure \ref{fig:LA_b} and table \ref{report1}. 

\subsection{Generalized axions}

In the above two models with broken translation symmetry, the mass square of
the graviton has only one term. One may wonder if Neural ODEs can learn more
general mass function. Here we will study the generalized axion (GA) model 
\cite{Baggioli2014}, which is a productive tool in the field of holographic
condensed matters \cite{Matteo2021}, see a recent application in amorphous
solids \cite{Li2108}. Of particular interest to us, it can assign more
general mass function to the graviton, while remaining analytically
tractable.

Let's generalize the bulk action (\ref{LA action}) to the form%
\begin{equation}
S_{\mathrm{bulk}}=\int d^{4}x\sqrt{-g}\left( \mathcal{R}+6-V(X)\right) .
\end{equation}%
We will select the function $V(X)$ that has been used in \cite{Baggioli2014} 
\begin{equation}
V(X)=\gamma _{1}X+\gamma _{5}X^{5},  \label{VX}
\end{equation}%
where $\gamma _{1}$ and $\gamma _{5}$ are two parameters.

Taking the variation of the action, we can write down the equations of motion%
\begin{eqnarray}
R_{\mu \nu }-\frac{1}{2}g_{\mu \nu }R-3g_{\mu \nu } &=& X_{\mu \nu }V^{\prime
}(X)-\frac{1}{2}g_{\mu \nu }V(X),   \\
\nabla _{\mu }V^{\prime }\nabla ^{\mu }\chi ^{I}+V^{\prime }\nabla ^{2}\chi
^{I} &=& 0.
\end{eqnarray}%
They have the analytical solution%
\begin{eqnarray}
f(z) &=& 1-z^{3}+\frac{1}{2}\sum_{n=1,5}\gamma _{n}\frac{z^{3}-z^{2n}}{3-2n},
\label{metricGA}    \nonumber  \\
\chi ^{1} &=& x,\;\chi ^{2}=y.
\end{eqnarray}%
From the metric, we can read the temperature%
\begin{equation}
T=\frac{1}{8\pi }\left( 6-\sum_{n=1,5}\gamma _{n}\right) .
\end{equation}%
We can also derive the graviton mass from eq. (\ref{m2}) and%
\begin{equation}
T_{\mu \nu }=2X_{\mu \nu }V^{\prime }(X)-g_{\mu \nu }V(X),
\end{equation}%
which yields%
\begin{equation}
m^{2}=\sum_{n=1,5}n\gamma _{n}z^{2n}.  \label{massGA}
\end{equation}

\begin{figure}[h]
\centering
\begin{subfigure}{0.326\textwidth}
    	\includegraphics[width=\textwidth]{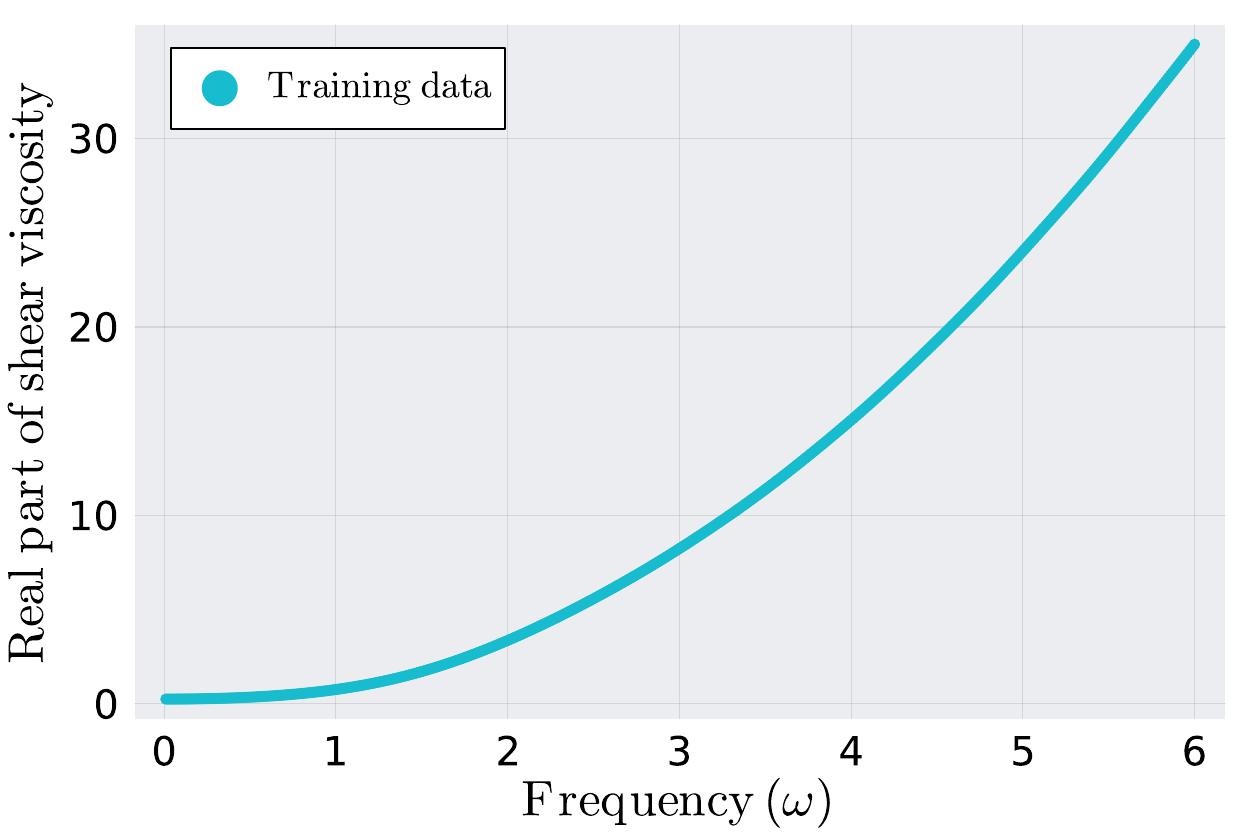}
  	\end{subfigure}
\begin{subfigure}{0.326\textwidth}
    	\includegraphics[width=\textwidth]{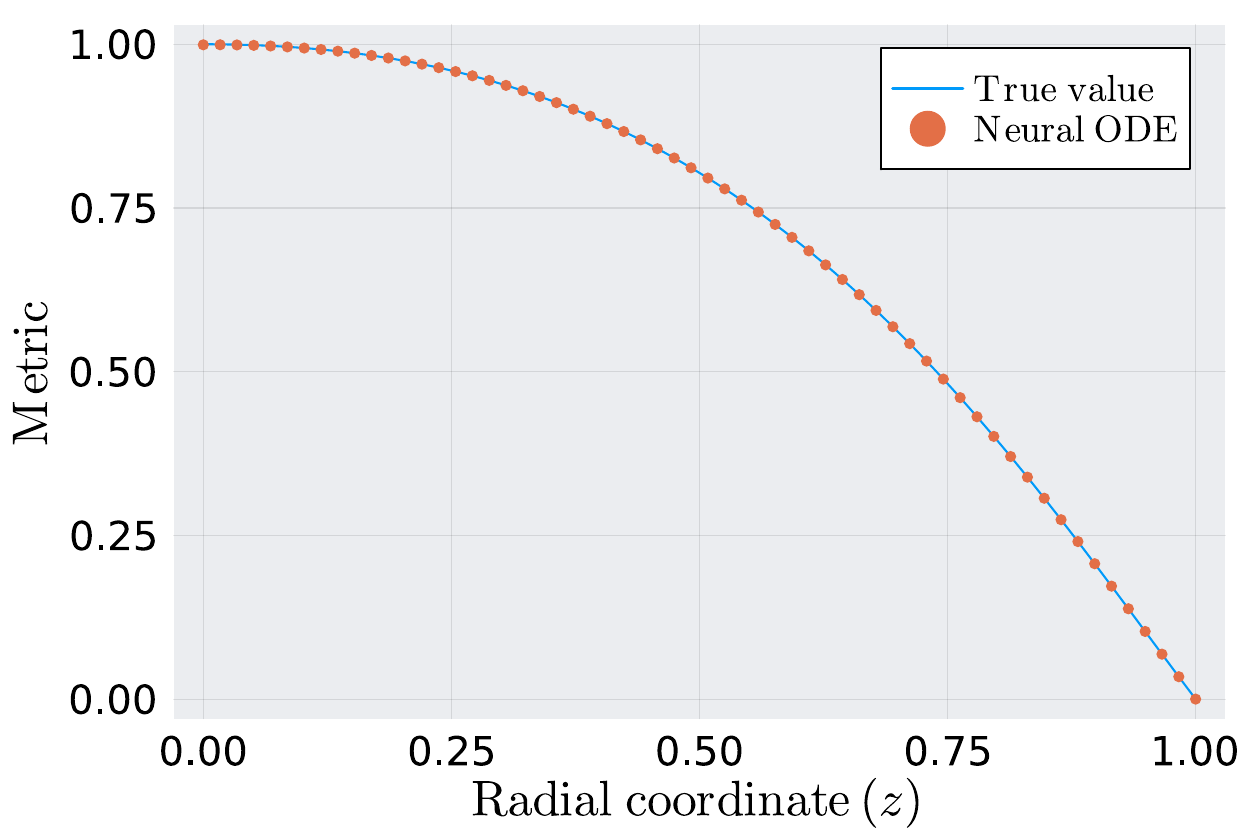}
  	\end{subfigure}
\begin{subfigure}{0.326\textwidth}
    	\includegraphics[width=\textwidth]{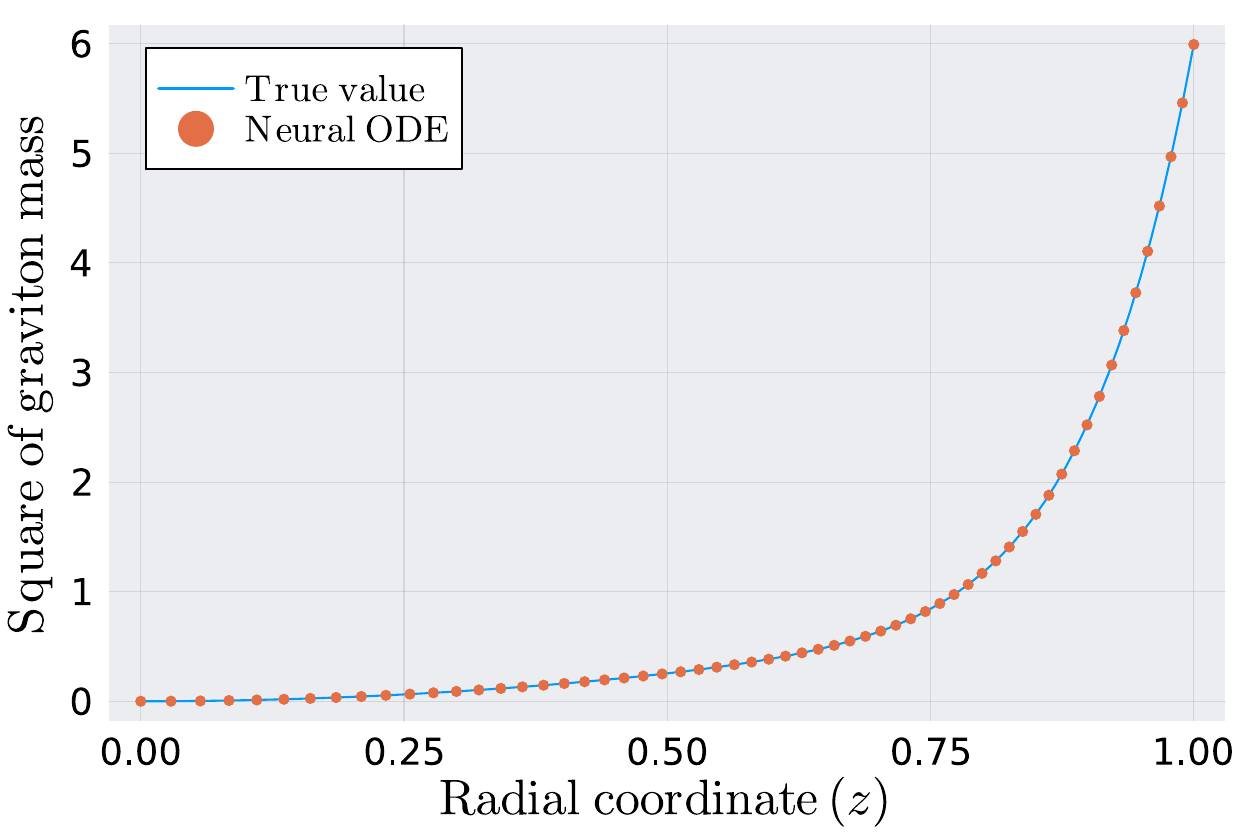}
  	\end{subfigure}
\caption{Training data and performance of Neural ODEs for the GA model.}
\label{fig:GA}
\end{figure}

Using eq. (\ref{metricGA}) and eq. (\ref{massGA}), the data are generated in
the left panel of figure \ref{fig:GA}. Here we have fixed the parameters $%
\gamma _{1}=\gamma _{5}=1$. We exhibit the training results in middle and
right panels of figure \ref{fig:GA}, as well as in figure \ref{fig:LA_b} and
table \ref{report1}. 

\section{Conclusion and discussion}

This work introduces an important physical ingredient to AdS/DL: translation
symmetry breaking. Specifically, we extend the data-driven holographic
transport model proposed in \cite{Wu2020}, allowing the machine to learn the
black hole metric or the graviton mass from the real part of the shear
viscosity\footnote{%
In \cite{Wu2020}, the model is subject to the several constraints: minimal
coupling of 3+1 dimensional Einstein gravity and matter, homogeneous and
isotropic background, vanishing graviton mass, and the relation $\eta =\chi
+i\omega r$ on the boundary. In the current work, we allow for a non-zero
graviton mass and reduce the boundary relation to its real part.}. In
addition, we use the Neural ODE to learn continuous functions, thereby
avoiding the problem of discretization errors. Around these physical and
technical results, we have made the following comments and explorations.

1) Universality of theoretical framework

Although some assumptions still remain, the current theoretical framework,
which mainly consists of the wave equation (\ref{wave eq0}), the regularity
at the horizon (\ref{chiIR}), and the boundary relation (\ref{UV relation}),
is very general. Considering that the holographic calculation of shear
viscosity does not impose the probe limit, this generality is quite rare and
precious.

2) Reconstruction of graviton mass

There are many ways to reconstruct the function of metric, but there has
been no way to reconstruct the function of graviton mass before. Our work
fills this gap, which may help to better understand the gravity dual of
translation symmetry breaking in bottom-up holographic models.

3) At low temperatures

In Table \ref{report1}, it is shown that the Neural ODE can achieve high
accuracy for the bulk reconstruction. However, we caution that this is
associated with the current model parameters ($\alpha ,\beta ,\gamma $),
which are\ simply fixed to 1. As these parameters increase, the temperature
of black holes decreases. Meanwhile, we observe a rapid degradation in the
performance of Neural ODEs\footnote{%
Similar phenomena were previously reported in \cite{Li:2022zjc}.}. To
address this issue, we will specify the temperature by hyperparameter tuning
and input several inductive biases. In this way, Neural ODEs can still
perform well, see Appendix C.

4) Entanglement entropy

In \cite{Hashimoto:2018bnb}, after the bulk metric is learned from lattice
QCD data, the quark-antiquark potential is calculated using the holographic
method. Interestingly, this potential exhibits both a linear confining part
and a Debye screening part. In Appendix D, we conduct a similar exploration.

\begin{figure}[h]
\centering
	\begin{subfigure}{0.5\textwidth}
    	\includegraphics[width=\textwidth]{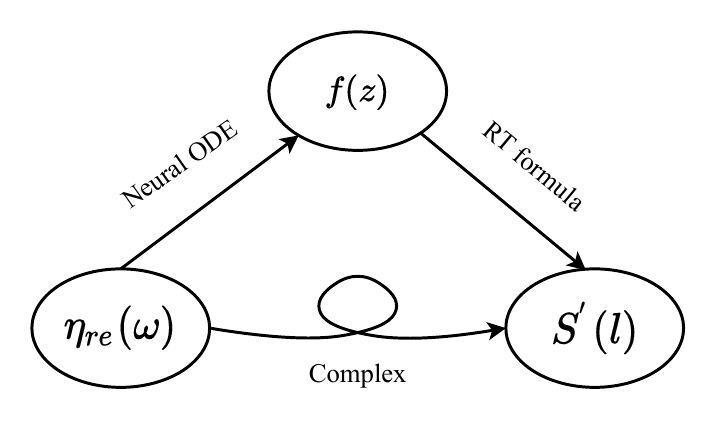}
  	\end{subfigure}
\caption{The schematic diagram for the relationship between $\eta _{\mathrm{%
re}}(\omega )$ and $S^{\prime }(l)$. At the bulk level, they are connected
by the Neural ODE and the RT formula through the metric, which is
represented by the straight arrows above. The curved arrow below indicates
that the relationship between them can be very complicated from the
perspective of field theory.}
\label{fig:eta_dS}
\end{figure}

Using the bulk metric $f(z)$ learned from the shear viscosity $\eta _{%
\mathrm{re}}(\omega )$, we calculate the derivative $S^{\prime }(l)$, where $%
S$ is the holographic EE and $l$ is the spatial size of entangling region%
\footnote{%
We focus on the derivative of EE rather than the EE itself because the
former is not sensitive to the UV cutoff \cite%
{Jokela:2020auu,Jokela2304,XWB2023}.}. Across a wide range of $l$ spanning
several orders of magnitude, we find that the MRE of $S^{\prime }(l)$ is
very small. This indicates that we have established a relationship between $%
S^{\prime }(l)$ and $\eta _{\mathrm{re}}(\omega )$. Note that this
relationship may be very complex from the perspective of field theories,
partly because $S^{\prime }(l)$ and $\eta _{\mathrm{re}}(\omega )$ belong to
two distinct classes of observables: quantum information measures and
transport coefficients. Nevertheless, using the bulk metric as a medium, $%
S^{\prime }(l)$ and $\eta _{\mathrm{re}}(\omega )$ can indeed be linked via
the Neural ODE and the Ryu-Takayanagi (RT) formula \cite{Ryu:2006bv}, see
figure \ref{fig:eta_dS}.

More importantly, the relationship between $S^{\prime }(l)$ and $\eta _{%
\mathrm{re}}(\omega )$ is independent with the probe limit and is not
confined to a specific model\footnote{%
For any effective methods of bulk metric reconstruction, their field-theory
observables are interrelated, reflecting at least a redundancy in the bulk
geometry information. However, each reconstruction method has its applicable
range. If these ranges are very limited or only slightly overlap, we would
not expect a clear relationship between the field-theory observables.}.
Based on its generality and robustness\footnote{%
It is found that the relative error of $S^{\prime }(l)$ decreases
oscillatingly as $l$ increases, indicating a more robust relationship in the IR. This robustness could facilitate the emergence of the relationship in the real-world strongly coupled systems that lack conformal symmetry in the UV.}, we conjecture that this relationship may manifest in
some real-world strongly coupled systems. In particular, given the recent
progress in lattice gauge theories with respect to both $S^{\prime }(l)$ and
the spectral function related to shear viscosity \cite%
{Jokela2304,Altenkort2211,Yao2402}, we expect that this conjecture could be
verified in the near future.

5) Beyond AdS/CFT

The current theoretical framework necessitates that the bulk spacetime be
asymptotically AdS. However, we deliberately refrain from explicitly
communicating this information to the machine. Our intentional omission
increases the training difficulty considerably, but the performance of
Neural ODEs remains excellent. This encouraging result suggests the potential to develop a
machine-learning framework that extends beyond asymptotic AdS, encompassing the Lifshitz and hyperscaling-violated geometries, as well as
other novel behaviors\footnote{%
By constructing a free bulk theory from the generalized free fields on the
boundary, it was found recently that the Sachdev-Ye-Kitaev model retains
geometric bulk features beyond its conformal limit \cite{Qi2306}. Notably,
the curvature of the emergent spacetime diverges near the boundary.}.

Ultimately, this work falls within the realm of reverse engineering of
AdS/CFT, where the holographic dictionary is given a priori. One can also
develop holographic machine learning that is independent of AdS/CFT \cite%
{You:2017guh,Hu:2019nea,Han:2019wue,Lam:2021ugb}, which explores the
emergence of spacetime and gravity from a more fundamental level.

\section*{Acknowledgments}

We thank Xian-Hui Ge, Yan Liu, Zhen-Kang Lu, Yu Tian, and Zhuo-Yu Xian for
helpful discussions. SFW was supported by NSFC grants (No.11675097,
No.12275166, and No. 12311540141).\bigskip

\newpage \appendix{} \setcounter{equation}{0} \renewcommand{\theequation}{A.%
\arabic{equation}}

\section{From shear response to shear viscosity}

Consider the bulk action of the Einstein gravity minimally coupled with the
matter%
\begin{equation}
S_{\mathrm{bulk}}=\int d^{4}x\sqrt{-g}\left( R+6+L_{\mathrm{matter}}\right) ,
\label{bulkEMA}
\end{equation}%
and the wave equation of the shear mode%
\begin{equation}
\frac{1}{\sqrt{-g}}\partial _{r}(\sqrt{-g}g^{rr}\partial _{r}h)+\left(
\omega ^{2}g^{tt}-m^{2}\right) h=0.  \label{wave eq}
\end{equation}%
We require that the metric and graviton mass can be expanded near the AdS
boundary, with the form%
\begin{eqnarray}
g_{tt} &=&r^{2}(1+\frac{a_{1}}{r}+\frac{a_{2}}{r^{2}}+\frac{a_{3}}{r^{3}}%
+\cdots ),  \notag \\
g^{rr} &=&r^{2}(1+\frac{b_{1}}{r}+\frac{b_{2}}{r^{2}}+\frac{b_{3}}{r^{3}}%
+\cdots ),  \notag \\
m^{2} &=&r^{-v}(c_{0}+\frac{c_{1}}{r}+\frac{c_{2}}{r^{2}}+\frac{c_{3}}{r^{3}}%
+\cdots ),  \label{m}
\end{eqnarray}%
where $\left( a_{i},b_{i},c_{i}\right) $ are some constants and $v$ is a
positive integer.

Inserting eq. (\ref{m}) into eq. (\ref{wave eq}), one can find that the
asymptotic solution of the wave equation can be written as%
\begin{equation}
h=h^{(0)}\left( 1+d_{1}\frac{1}{r^{1}}+d_{2}\frac{1}{r^{2}}+d_{3}\frac{\log
(r)}{r^{3}}\right) +h^{(3)}\frac{1}{r^{3}}+\cdots .  \label{h}
\end{equation}%
Here $h^{(0)}$ is the source, $h^{(3)}$ depends on $h^{(0)}$ and the
incoming boundary condition on the horizon, and $d_{i}$ is determined by the
frequency and the constants $a_{i}$, $b_{i}$, $c_{i}$ and $v$.

To ensure a well-defined variational system, we need\ the Gibbons-Hawking
term and the counterterms%
\begin{equation}
S_{\mathrm{GH}}=2\int d^{3}x\sqrt{-\gamma }K,\;S_{\mathrm{ct}}=\int d^{3}x%
\sqrt{-\gamma }\left( -4-R+L_{\mathrm{matter}}^{(1)}\right) ,  \label{GHEMA}
\end{equation}%
where $L_{\mathrm{matter}}^{(1)}$ denotes the contribution of the
counterterms from the matter. We will expand the on-shell bulk action, the
Gibbons-Hawking term, and the counterterms to the second order of the shear
mode. Here the key step is to calculate the second-order\ variation of the
matter action $S_{\mathrm{m}}=\int d^{4}x\sqrt{-g}L_{\mathrm{matter}}$ as
follows:%
\begin{eqnarray}
\frac{\delta ^{2}S_{\mathrm{m}}}{\delta g_{xy}^{2}} &=&\frac{1}{2}\left[ 
\frac{\delta S_{\mathrm{m}}}{\delta g^{xx}}\frac{\delta ^{2}g^{xx}}{\left(
\delta g_{xy}\right) ^{2}}+\frac{\delta ^{2}S_{\mathrm{m}}}{\left( \delta
g^{xx}\right) ^{2}}\left( \frac{\delta g^{xx}}{\delta g_{xy}}\right)
^{2}+x\leftrightarrow y\right]  \notag \\
&&+\frac{1}{2}\left[ \frac{\delta S_{\mathrm{m}}}{\delta g^{xy}}\frac{\delta
^{2}g^{xy}}{\left( \delta g_{xy}\right) ^{2}}+\frac{\delta ^{2}S_{\mathrm{m}}%
}{\left( \delta g^{xy}\right) ^{2}}\left( \frac{\delta g^{xy}}{\delta g_{xy}}%
\right) ^{2}+x\leftrightarrow y\right] .
\end{eqnarray}%
Since $T_{\mu \nu }=-2\frac{1}{\sqrt{-g}}\frac{\delta S_{\mathrm{m}}}{\delta
g^{\mu \nu }}$ and%
\begin{equation}
g^{xx}=g^{yy}=\frac{g_{xx}}{-g_{xy}^{2}+g_{xx}^{2}},\;g^{xy}=\frac{-g_{xy}}{%
-g_{xy}^{2}+g_{xx}^{2}},
\end{equation}%
one can read%
\begin{equation}
\frac{\delta ^{2}S_{\mathrm{m}}}{\delta g_{xy}^{2}}=-\frac{\sqrt{-g}}{%
g_{xx}^{2}}\left( \frac{1}{g_{xx}}T_{xx}-\frac{1}{2}\frac{\delta T_{xy}}{%
\delta g_{xy}}\right) .  \label{dSdg2}
\end{equation}

Using eq. (\ref{dSdg2}) and the Einstein equation for the background metric%
\begin{equation}
\frac{1}{2}T_{xx}=-3g_{xx}-\frac{g_{xx}g_{rr}^{\prime }g_{tt}^{\prime }}{%
4g_{rr}^{2}g_{tt}}-\frac{g_{xx}g_{tt}^{\prime 2}}{4g_{rr}g_{tt}^{2}}-\frac{%
g_{rr}^{\prime }g_{xx}^{\prime }}{4g_{rr}^{2}}+\frac{g_{tt}^{\prime
}g_{xx}^{\prime }}{4g_{rr}g_{tt}}-\frac{g_{xx}^{\prime 2}}{4g_{rr}g_{xx}}+%
\frac{g_{xx}g_{tt}^{\prime \prime }}{2g_{rr}g_{tt}}+\frac{g_{xx}^{\prime
\prime }}{2g_{rr}},
\end{equation}%
we can perform the expansion and express the result in the frequency space%
\begin{equation}
\left. S_{\mathrm{bulk}}+S_{\mathrm{GH}}+S_{\mathrm{ct}}\right\vert _{%
\mathrm{on-shell}}=\int d^{2}x\int_{-\infty }^{\infty }\frac{d\omega }{2\pi }%
\frac{1}{2}\left[ \left( L_{\mathrm{gravity}}^{(2)}+L_{\mathrm{matter}%
}^{(2)}\right) \bar{h}h-g_{xx}\sqrt{\frac{g_{tt}}{g_{rr}}}\bar{h}h^{\prime }%
\right] ,  \label{qaction}
\end{equation}%
where $\bar{h}$ has the argument $-\omega $. In the terms $\sim \bar{h}h$, $%
L_{\mathrm{gravity}}^{(2)}$ can be referred as the gravity contribution and $%
L_{\mathrm{matter}}^{(2)}$ denotes the contribution from $L_{\mathrm{matter}%
}^{(1)}$. One can find that $L_{\mathrm{gravity}}^{(2)}$ has the form 
\begin{equation}
L_{\mathrm{gravity}}^{(2)}=-\omega ^{2}\frac{g_{xx}}{\sqrt{g_{tt}}}+4\sqrt{%
g_{tt}}g_{xx}-\frac{g_{xx}g_{tt}^{\prime }}{\sqrt{g_{tt}g_{rr}}}-\sqrt{\frac{%
g_{tt}}{g_{rr}}}g_{xx}^{\prime }.  \label{L2g}
\end{equation}%
Interestingly, it is independent with the energy-momentum tensor and the
gravition mass. In contrast, we can not write down the general form of $L_{%
\mathrm{matter}}^{(2)}$.

One may notice that $L_{\mathrm{gravity}}^{(2)}$ and $L_{\mathrm{matter}%
}^{(2)}$ usually diverge at the boundary. To deal with the divergence, we
input eq. (\ref{m}) and eq. (\ref{h}) into eq. (\ref{qaction}), which yields
the renormalized action%
\begin{equation}
S_{\mathrm{ren}}=\int d^{2}x\int_{-\infty }^{\infty }\frac{d\omega }{2\pi }%
\frac{1}{2}\left[ \left( L_{\mathrm{gravity}}^{(3)}+L_{\mathrm{matter}%
}^{(3)}\right) \bar{h}^{(0)}h^{(0)}+3\bar{h}^{(0)}h^{(3)}\right] ,   \label{Sren}
\end{equation}%
where $L_{\mathrm{gravity}}^{(3)}$ and $L_{\mathrm{matter}}^{(3)}$ are
finite and they are contributed by the gravity and the matter, respectively.
In terms of eq. (\ref{L2g}), it is obvious that $L_{\mathrm{gravity}}^{(3)}$
is real since it is composed of the real constants in eq. (\ref{m}).
Moreover, we will assume that $L_{\mathrm{matter}}^{(3)}$ is also real. This
assumption is valid for all holographic models in the main text. To prove
this, we will list their counterterms explicitly.

1) Massive gravity

The counterterms of massive gravity have been derived in \cite{Wu1903}. They
can be expressed as%
\begin{equation}
S_{\mathrm{ct}}=\int d^{3}x\sqrt{-\gamma }\left[ -4-R+\frac{1}{2}\alpha
e_{1}+\frac{1}{16}\alpha ^{2}e_{2}+\frac{\alpha }{4}(2R_{ij}Y^{ij}-e_{1}R)%
\log r\right] ,  \label{CTMG}
\end{equation}%
from which we calculate%
\begin{equation}
L_{\mathrm{matter}}^{(2)}=-\frac{1}{8}\alpha \left( \sqrt{g_{tt}g_{xx}}%
+2\omega ^{2}\sqrt{\frac{g_{xx}}{g_{tt}}}\log r\right) .  \label{Lm2MG}
\end{equation}

2) Linear axions

The counterterms of the linear axion model are simple:%
\begin{equation}
S_{\mathrm{ct}}=\int d^{3}x\sqrt{-\gamma }\left( -4-R+\gamma
^{ij}X_{ij}\right) .  \label{CTLA}
\end{equation}%
We can read%
\begin{equation}
L_{\mathrm{matter}}^{(2)}=-\frac{1}{2}\beta ^{2}\sqrt{g_{tt}}.  \label{Lm2LA}
\end{equation}

3) Generalized axions

In terms of the holographic renormalization studied in \cite{Ma2208}, one
can infer that the counterterms of generalized axion model are given by%
\begin{equation}
S_{\mathrm{ct}}=\int d^{3}x\sqrt{-\gamma }\left( -4-R+\gamma _{1}\gamma
^{ij}X_{ij}\right) ,  \label{CTGA}
\end{equation}%
which indicates%
\begin{equation}
L_{\mathrm{matter}}^{(2)}=-\frac{1}{2}\gamma _{1}\sqrt{g_{tt}}.
\label{Lm2GA}
\end{equation}%
From eq. (\ref{Lm2MG}), eq. (\ref{Lm2LA}), and eq. (\ref{Lm2GA}), it is
obvious that they can only generate real $L_{\mathrm{matter}}^{(3)}$.

Motivated by these explicit examples, let's move on a more general analysis.
Suppose that a counterterm is an intrinsic scalar on the boundary, which
consists of the induced metric, matter fields, and their derivatives. To be
safe, we will focus on neutral matter fields and impose the parity symmetry%
\footnote{%
Then the Levi-Civita tensor can be excluded in building the counterterm.}.
One can see that $L_{\mathrm{matter}}^{(3)}$ cannot have an imaginary part
unless the expansion of $L_{\mathrm{matter}}^{(1)}$ contains a term%
\begin{equation}
L_{\mathrm{im}}\sim \gamma ^{tt}\partial _{t}\gamma _{xy}\gamma _{xy}\left(
\Psi ,\partial \Psi \right) _{t\cdots },  \label{Lim}
\end{equation}%
where $\left( \Psi ,\partial \Psi \right) _{t\cdots }$ denotes certain
matter field or its spatial derivative with the indices $t\cdots $. The
expression of $L_{\mathrm{im}}$ is understood that the subscript $t$ in $%
\partial _{t}\gamma _{xy}$ must be paired with the subscript $t$ in $\left(
\Psi ,\partial \Psi \right) _{t\cdots }$ in order to contract with the
superscripts of $g^{tt}$. Obviously, $\Psi $ is neither a scalar nor a
spinor since they have no spacetime indices. $\Psi $ is also not a U(1)
vector: Due to the gauge symmetry, the U(1) vector can only appear in the
form of $F_{ij}F^{ij}$, which cannot be paired with $\partial _{t}\gamma
_{xy}$. In light of these analysis, one can find that our assumption is
valid for all Einstein-Maxwell-Dirac-scalar models with parity symmetry and
neutral matters. For more general models, we cannot provide a similar proof
by checking eq. (\ref{Lim}). Nevertheless, keeping in mind the intrinsic
structure of counterterms\footnote{%
Even if one can construct a counterterm ansatz that yields $L_{\mathrm{im}}$%
, this does not necessarily imply that it is a genuine counterterm for a
specific model. In \cite{Ma2208}, it was found that the exact ansatz of
counterterms can be generated from the radial Hamiltonian, indicating the
existence of an intrinsic structure of counterterms.}, it is reasonable to
suspect that the current assumption may be valid in a more general situation%
\footnote{%
For example, consider the famous model of holographic superconductor \cite%
{Bradlyn0803}. The charged complex scalar field $\Phi $ is coupled with the
U(1) vector $A$, which in principle might produce: $L_{\mathrm{im}}\sim
\gamma ^{tt}\partial _{t}\gamma _{xy}\gamma _{xy}\left\vert (\partial
_{t}-iA_{t})\Phi \right\vert $. However, the matter contribution to the
counterterm of this model is just $L_{\mathrm{matter}}^{(1)}\sim \left\vert
\Phi \right\vert ^{2} $.}. In fact, we are not aware of any obvious
counterexamples.

To proceed, we consider a field theory lived in a two-dimensional flat space
and define the retarded correlator\cite{Wu2020,Wu2015}%
\begin{equation}
G_{T^{xy}T^{xy}}^{R}\left( t\right) \equiv \frac{4\delta ^{2}W}{\delta h_{%
\mathrm{a}xy}\left( t\right) \delta h_{\mathrm{r}xy}\left( 0\right) },
\end{equation}%
where $W$ is the generating functional, $\delta h_{xy}$ denotes the shear
perturbation of the metric, and the subscripts $\left( \mathrm{r,a}\right) $
indicate the closed time-path formalism.

Using the holographic dictionary, one can read the retarded correlator from
eq. (\ref{Sren}): 
\begin{equation}
G_{T^{xy}T^{xy}}^{R}\left( \omega \right) =L^{(3)}+3\frac{h^{(3)}}{h^{(0)}},
\label{Gra}
\end{equation}%
where $L^{(3)}=L_{\mathrm{gravity}}^{(3)}+L_{\mathrm{matter}}^{(3)}$. Expand
the response function near the boundary, which yields%
\begin{equation}
\left. \chi (\omega )\right\vert _{r\rightarrow \infty }=\left. \frac{\Pi }{%
i\omega h}\right\vert _{r\rightarrow \infty }=\left. -\frac{\sqrt{-g}%
g^{rr}\partial _{r}h}{i\omega h}\right\vert _{r\rightarrow \infty }=\frac{1}{%
i\omega }\left. \left( \frac{3h^{(3)}}{h^{(0)}}-e_{1}\right) \right\vert
_{r\rightarrow \infty }.  \label{chi}
\end{equation}%
Here $e_{1}$ is real, relying on $r$, $\omega $, and previous constants.
Comparing eq. (\ref{Gra}) and eq. (\ref{chi}), we have%
\begin{equation}
\left. \chi (\omega )\right\vert _{r\rightarrow \infty }=\frac{1}{i\omega }%
G_{T^{xy}T^{xy}}^{R}(\omega )-\frac{1}{i\omega }\left. \left(
L^{(3)}+e_{1}\right) \right\vert _{r\rightarrow \infty }.  \label{xG}
\end{equation}%
Importantly, the last term of eq. (\ref{xG}), which depends on the UV
details of specific models, is purely imaginary.

In view of this, we will consider the Kubo formula for the\ real part of the
frequency-dependent shear viscosity \cite{Read1207,Wu2015,Wu2020}%
\begin{equation}
\eta _{\mathrm{re}}(\omega )=\frac{1}{\omega }\mathrm{Im}\left[
G_{T^{xy}T^{xy}}^{R}(\omega )\right] .  \label{Kubo}
\end{equation}%
Combining eq. (\ref{xG}) and eq. (\ref{Kubo}), we finally obtain a simple
equality between the shear viscosity and the shear response%
\begin{equation}
\eta _{\mathrm{re}}(\omega )=\left. \chi _{\mathrm{re}}(\omega )\right\vert
_{r\rightarrow \infty }.  \label{etachim}
\end{equation}

\section{Training scheme and report}

In order to perform training, we need to make some settings.

1) Neural network

We use a neural network to represent the metric or mass. It is a fully
connected feed-forward network consisting of three dense layers:%
\begin{equation}
(1,5)\rightarrow (5,5,\mathrm{tanh})\rightarrow (5,1).
\end{equation}%
For each layer, we have specified the number of input nodes and output
nodes. In the second layer, a tanh activation is used. Note that the neural
network has 46 trainable parameters, which include weights and biases.

2) ODE solver

Our Neural ODEs are solved using the Tsitouras 5/4 method \cite{Tsitouras},
which is an adaptive and explicit Runge-Kutta method.

3) IR and UV cutoffs

In solving the ODEs numerically, the horizon and boundary cannot be touched
exactly. We set the IR and UV cutoffs as $z_{\mathrm{IR}}=0.9999$ and $z_{%
\mathrm{UV}}=0.0001$.

4) Initial values

To learn the metric, the neural network is initialized by sampling from a
standard normal distribution $\mathcal{N}(0,1)$. The normal distribution $%
\mathcal{N}(0,0.001)$ with a small standard deviation is utilized for
learning the mass and hyperparameter tuning.

5) Hyperparameter tuning

\begin{figure}[h]
\centering
\begin{subfigure}{0.326\textwidth}
    	\includegraphics[width=\textwidth]{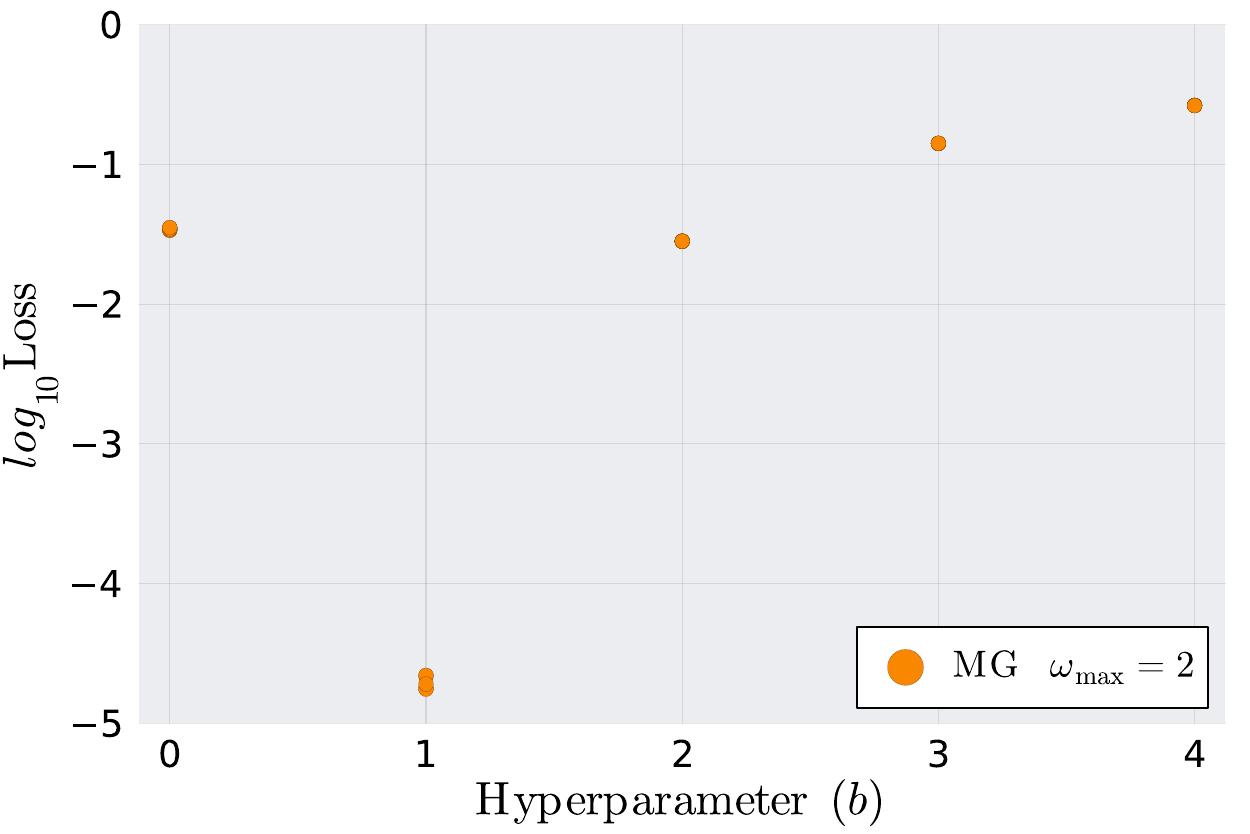}
  	\end{subfigure}
\begin{subfigure}{0.326\textwidth}
    	\includegraphics[width=\textwidth]{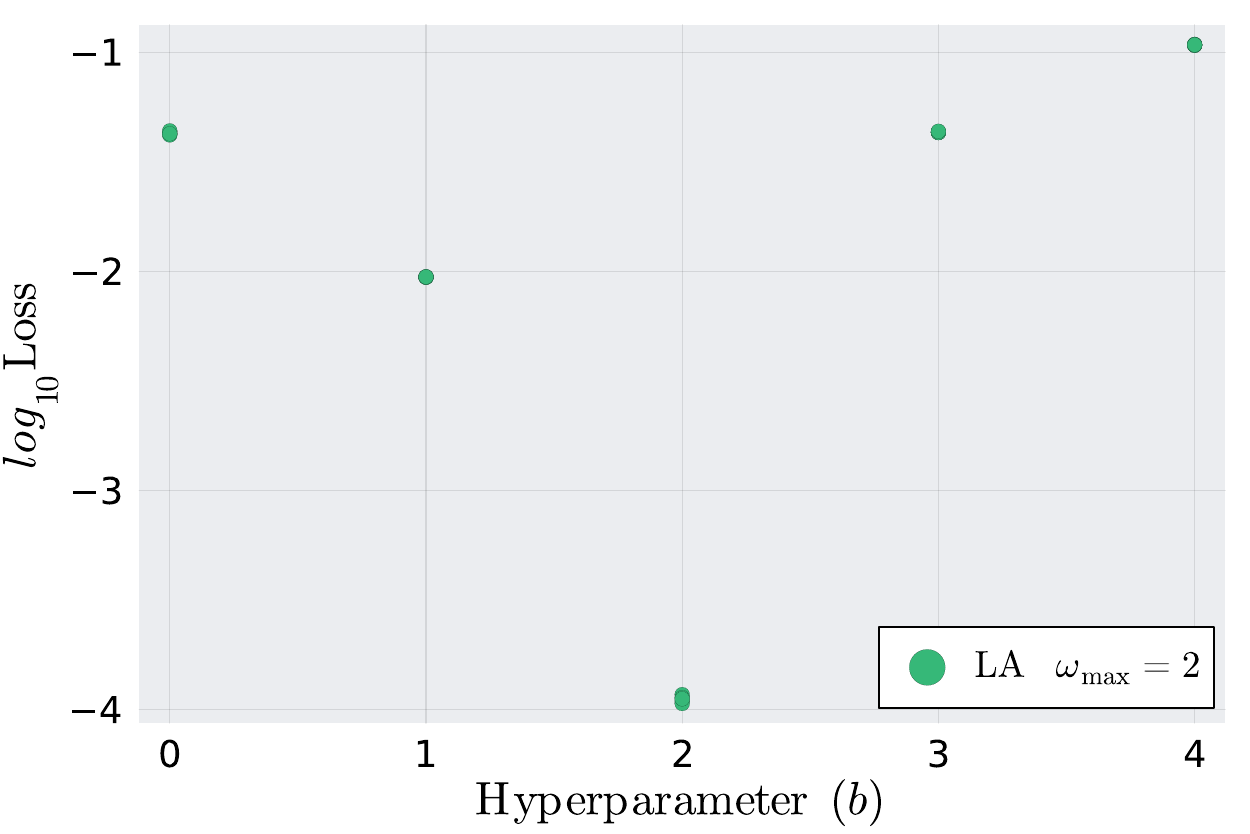}
  	\end{subfigure}
\begin{subfigure}{0.326\textwidth}
    	\includegraphics[width=\textwidth]{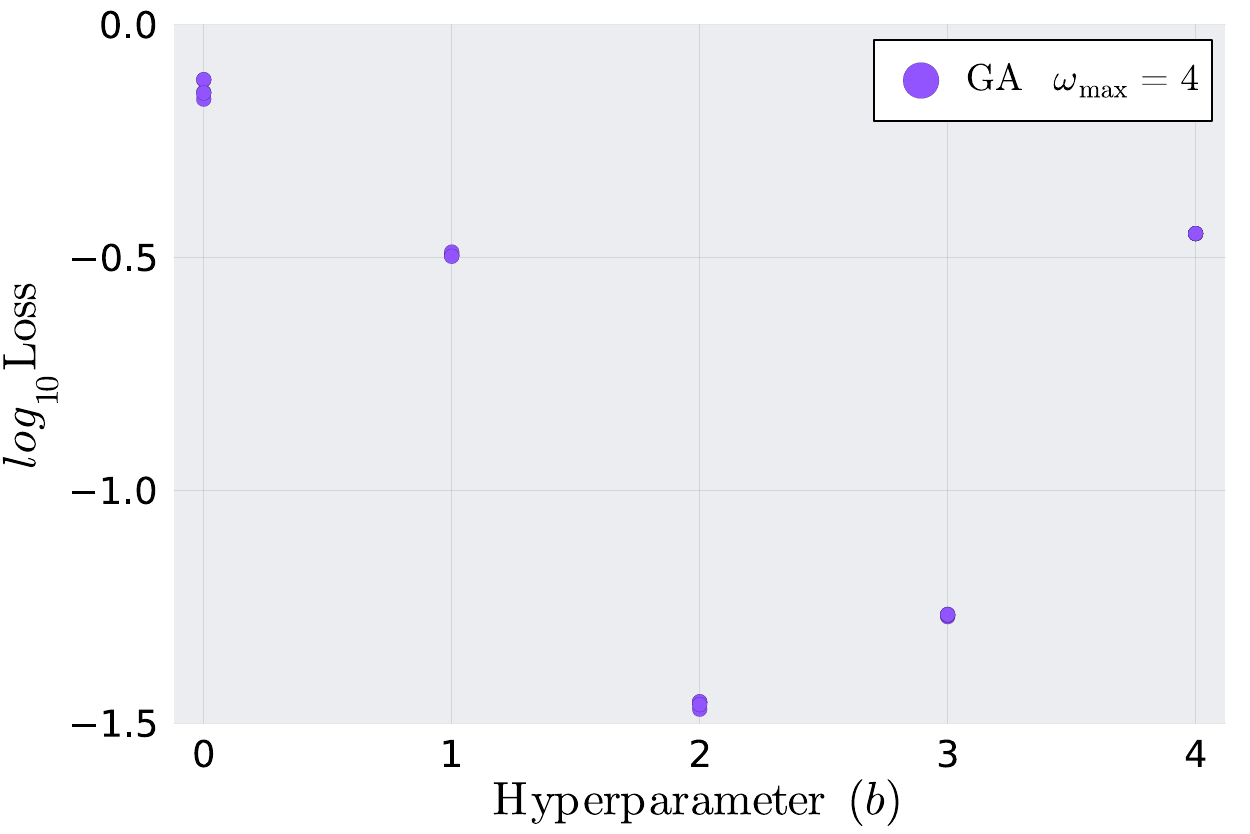}
  	\end{subfigure}
\caption{The hyperparameter tuning of $b$ in the MG (left), LA (middle) and
GA (right) models. Each $b$ is trained five\ times, yielding five nearly
coincident loss function points in most cases. As one can see, the optimal
hyperparameters for three models are 1, 2, and 2, respectively.}
\label{fig:LA_b}
\end{figure}

There are various hyperparameter tuning methods, which consume different
computing resources. Here we propose a simple way to quickly find the
exponent $b$ in the square of graviton mass. We define the search range as $%
[0,4]$ with the step size 1. We use the standard cross-validation method
with the simple hold-out technique \cite{Arlot}. We separate the dataset
within the frequency range $[0.01,\omega _{\max }]$ into two parts. The part
with $[0.01,\omega _{\max }-1]$ goes to the training set and the part with $%
[\omega _{\max }-0.99,\omega _{\max }]$ to the validation set. The rationale
for this separation lies in the dependence of the exponent $b$ on the UV
physics. We use the optimizers RMSProp and Adam in order. Their learning
rates, epoch numbers, and batch sizes are set as (0.001, 0.0001), (10, 10)
and (1, 1). We also add the L1 regularization term $L_{1}=0.1\times
\left\vert \theta \right\vert $ in the loss function. For each $b$ in the
search range, we train 5 times and take the average validation error to
select the optimal $b$. As long as $\omega _{\max }$ is large enough, the
correct $b$ can be found for the current three targets. Specifically, $%
\omega _{\max }=2$ is large enough for the MG and LA models, and $\omega
_{\max }=4$ is large enough for the GA model, see figure \ref{fig:LA_b}.

It should be noted that our hyperparameter tuning method has its
limitations. Considering the true mass square with the UV behavior $%
m^{2}=az^{b}+\cdots $, we observe that the required $\omega _{\max }$
increase as $a$ decreases. However, even if $b$ cannot be found accurately,
the final learned curve of the mass square usually deviates only slightly at
UV. In the future, more sophisticated hyperparameter tuning methods can be
explored to improve the results, if necessary.

6) Training process

The training process can be divided into multiple stages, which are
described as follows.

a) In most cases, the optimizers (RMSProp, Adam, Adam, BFGS) are used
sequentially in four stages. For learning the mass of the MG and the LA
models, only the first two training stages are needed, as the log-cosh loss
becomes negligible (less than the machine epsilon $\sim 1\times 10^{-16}$).

b) The learning rates (0.001, 0.0001, 0.00001) are used sequentially in
first three stages. The epoch numbers are (10, 10, 100) respectively. The
batch sizes are all set as 1. In the fourth stage, the initial size of the
step is set as 0.01 and the maximum number of iterations is set as 500.

c) In the first stage, we train $N_{1}=120$ times, each time with randomly
initialized trainable parameters. In the $i$-th stage with $i>1$, we obtain
the initial parameters from the top $N_{i}=N_{i-1}/2$ trained results with
the minimum loss in the $\left( i-1\right) $-th stage. We conduct parallel
training on multiple computer cores. When more than $N_{i}/2$ tasks are
completed, we stop the $i$-th stage and move on to the $\left( i+1\right) $%
-th stage. This early stopping can effectively prevent the machine from
getting stuck in solving Neural ODEs with bad parameters.

After training, we collect information on the best result, including the
minimum loss, the MRE of the learned metric, and the MRE of the learned mass
square, see table \ref{report1}. Note that the MRE is calculated by
uniformly sampling 100 points within the allowed range of the radial
coordinate $z$.

\begin{table}[h]
\centering%
\begin{tabular}{lllllll}
\hline\hline
Target & $f$ of MG & $m^{2}$ of MG & $f$ of LA & $m^{2}$ of LA & $f$ of GA & 
$m^{2}$ of GA \\ \hline
Loss & $7\times 10^{-10}$ & $<1\times 10^{-16}$ & $2\times 10^{-10}$ & $%
<1\times 10^{-16}$ & $6\times 10^{-10}$ & $1\times 10^{-11}$ \\ 
MRE & $5\times 10^{-5}$ & $3\times 10^{-9}$ & $1\times 10^{-5}$ & $2\times
10^{-10}$ & $7\times 10^{-5}$ & $6\times 10^{-4}$ \\ \hline\hline
\end{tabular}%
\caption{Minimum loss and MRE of six machine learning experiments. Their
learning targets are the metric and mass square in the models of MG, LA and
GA.}
\label{report1}
\end{table}

\section{At low temperatures}

\begin{figure}[h]
\centering
\begin{subfigure}{0.326\textwidth}
    	\includegraphics[width=\textwidth]{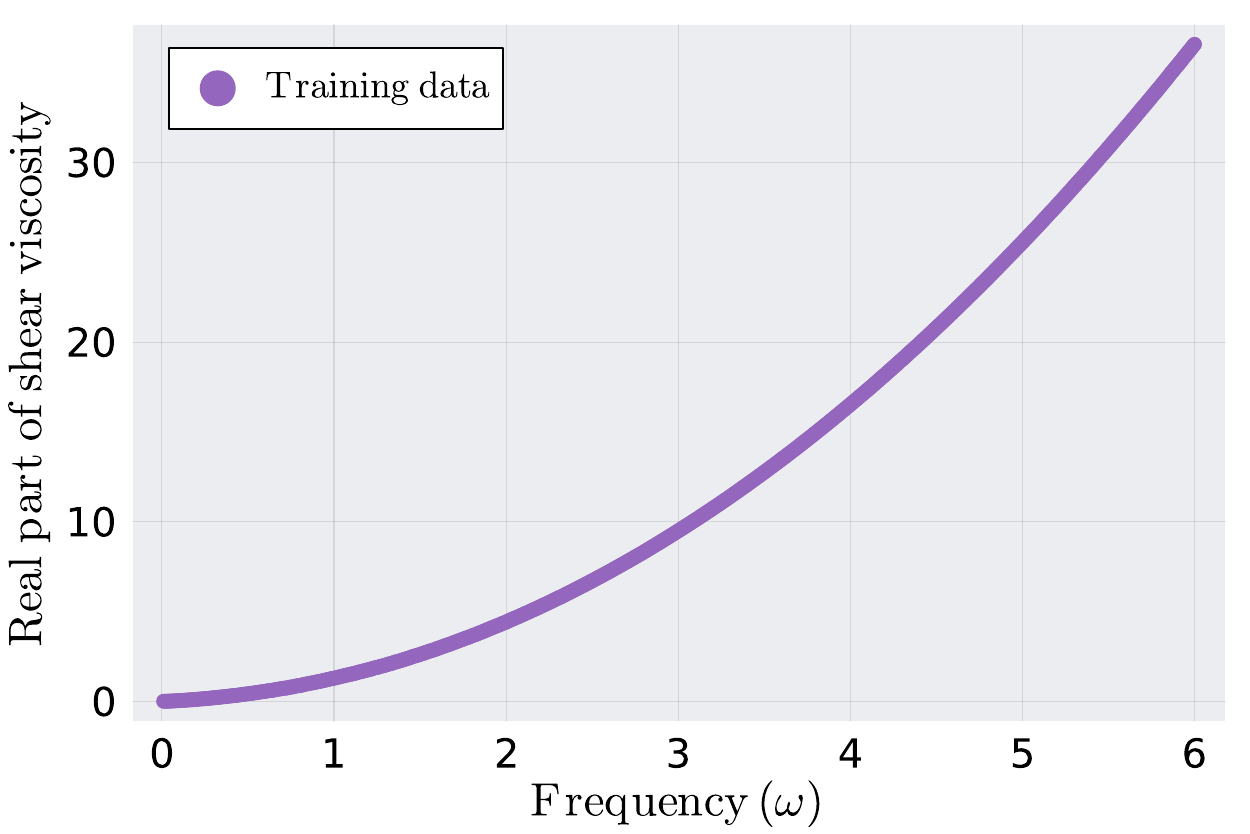}
  	\end{subfigure}
\begin{subfigure}{0.326\textwidth}
    	\includegraphics[width=\textwidth]{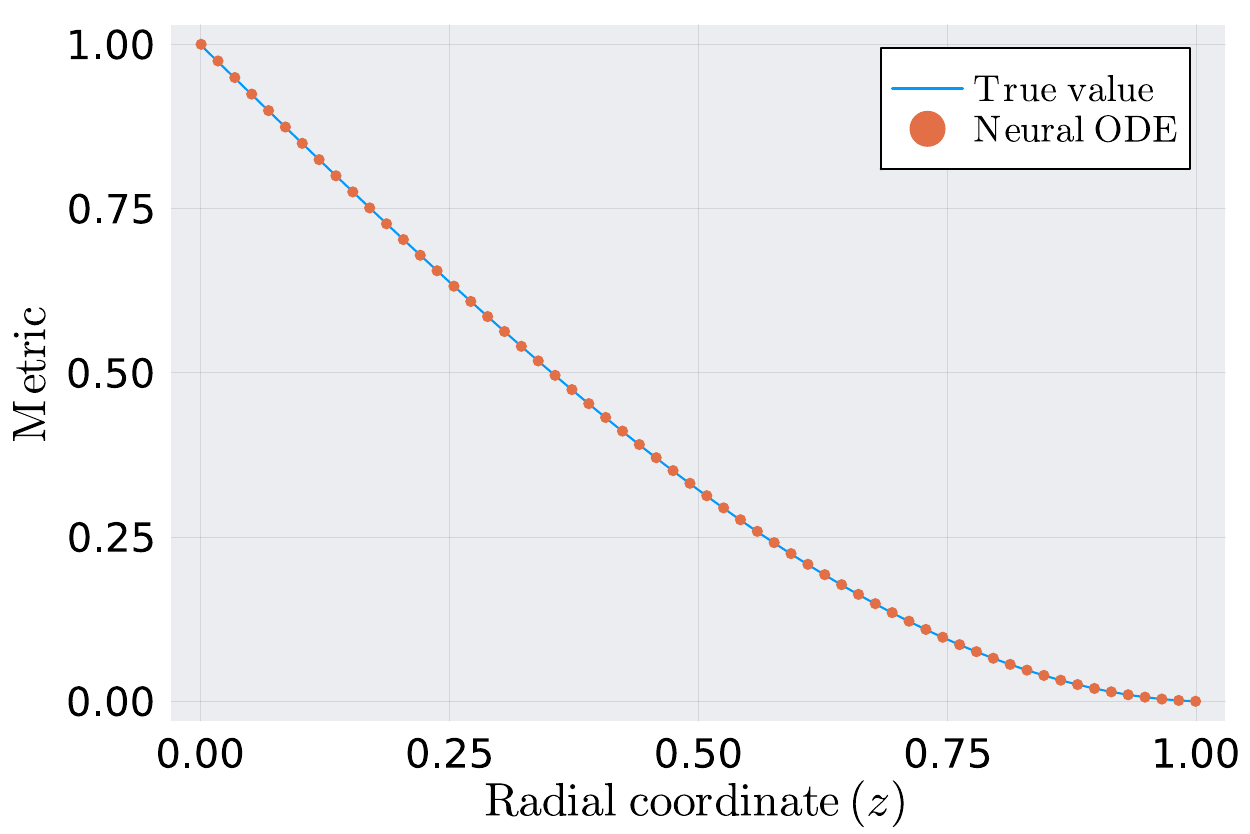}
  	\end{subfigure}
\begin{subfigure}{0.326\textwidth}
    	\includegraphics[width=\textwidth]{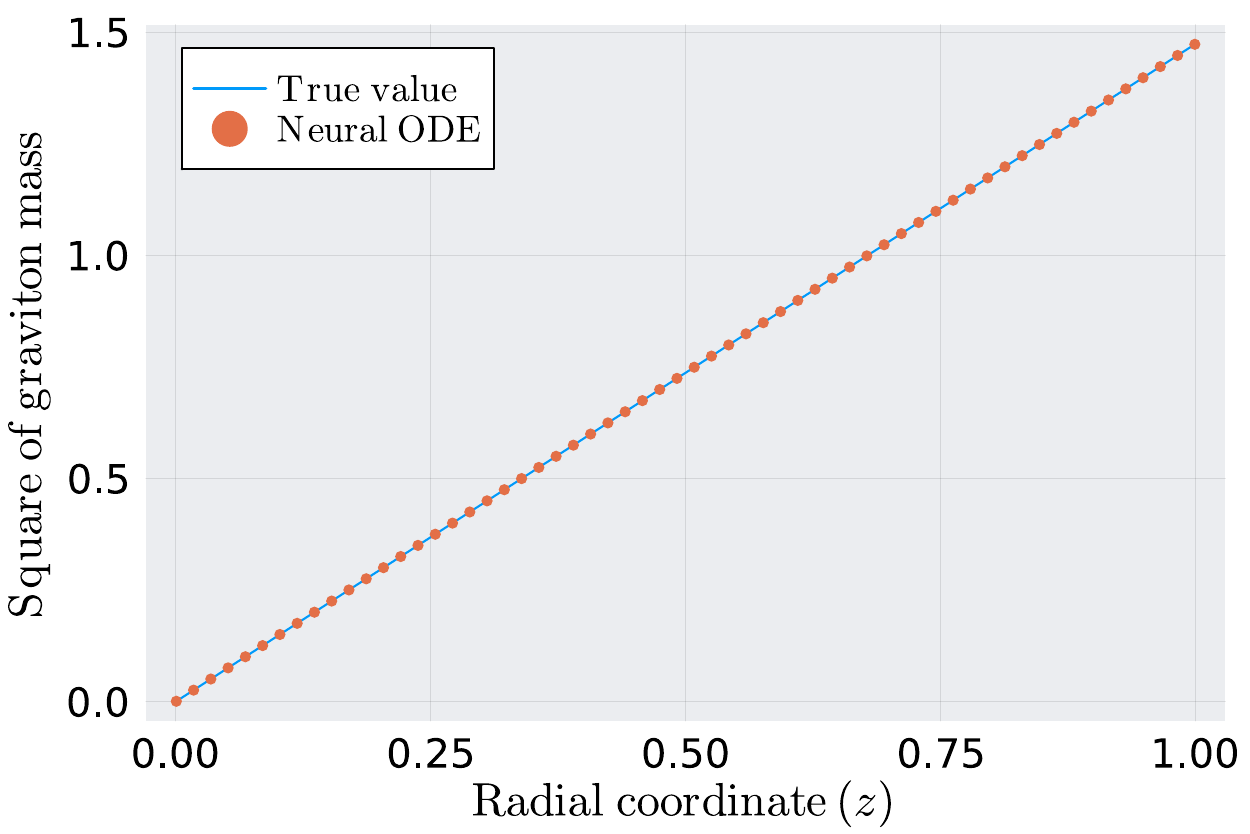}
  	\end{subfigure}   \\
\begin{subfigure}{0.326\textwidth}
    	\includegraphics[width=\textwidth]{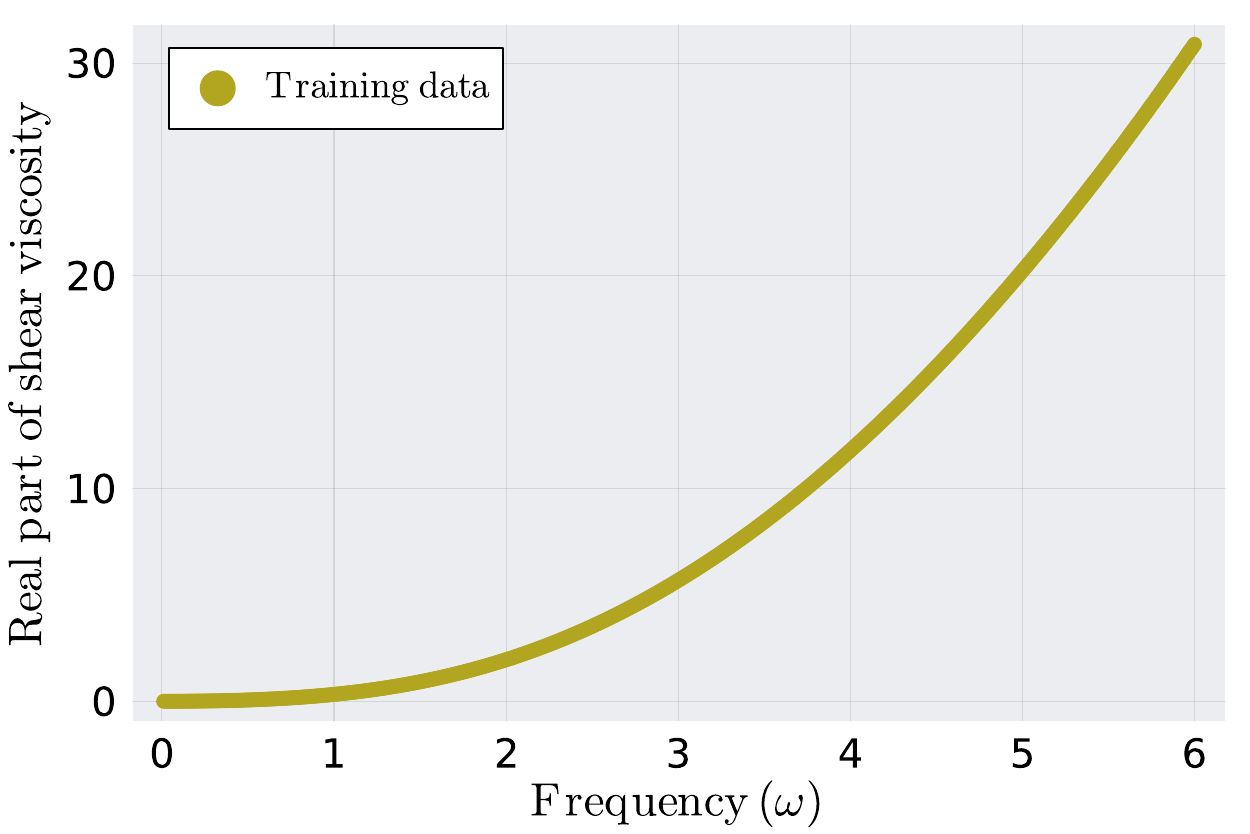}
  	\end{subfigure}
\begin{subfigure}{0.326\textwidth}
    	\includegraphics[width=\textwidth]{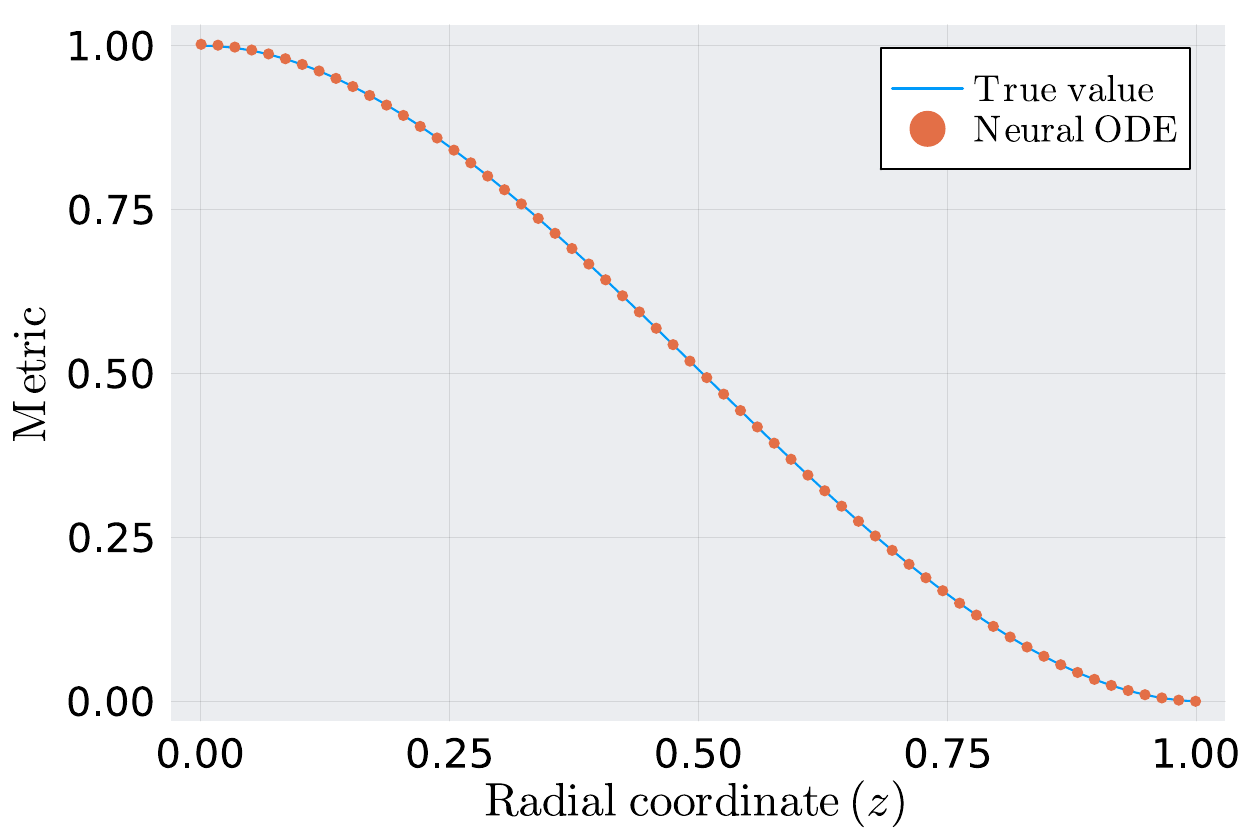}
  	\end{subfigure}
\begin{subfigure}{0.326\textwidth}
    	\includegraphics[width=\textwidth]{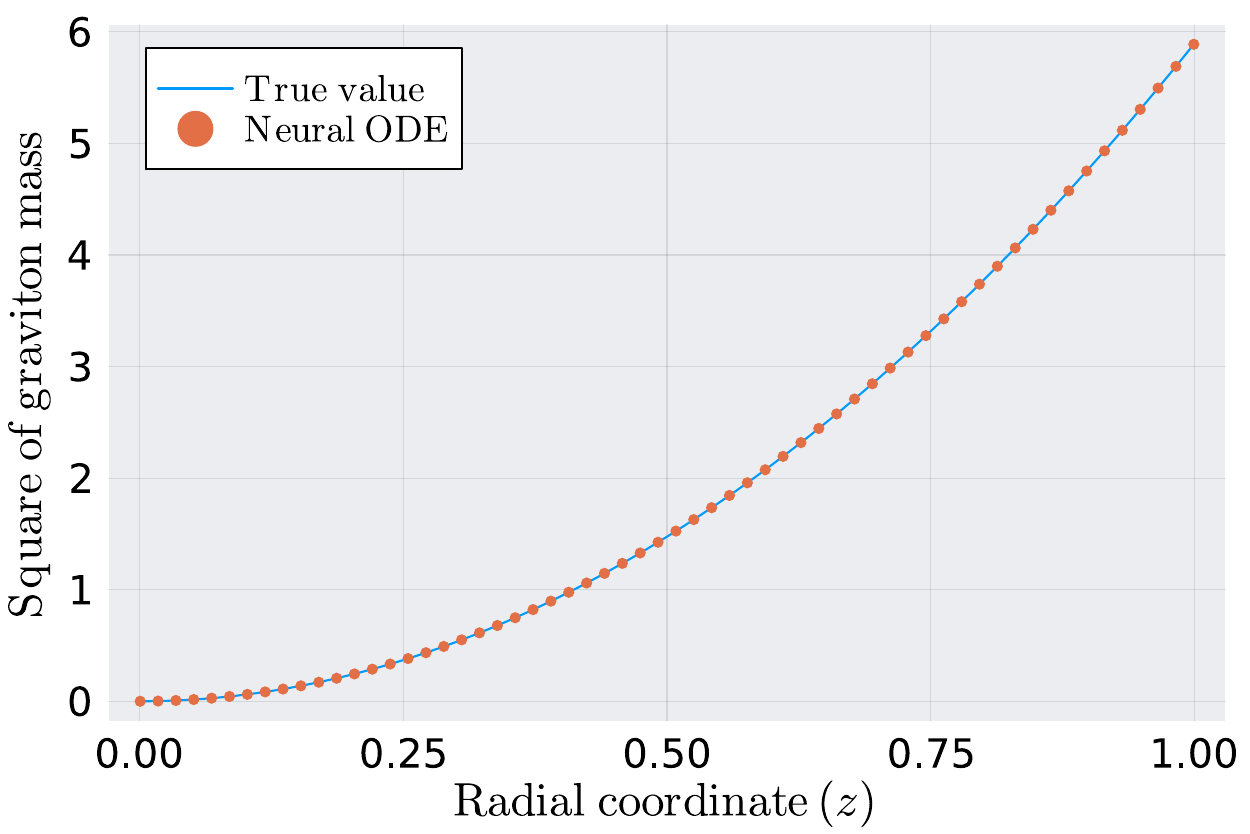}
  	\end{subfigure}  \\
\begin{subfigure}{0.326\textwidth}
    	\includegraphics[width=\textwidth]{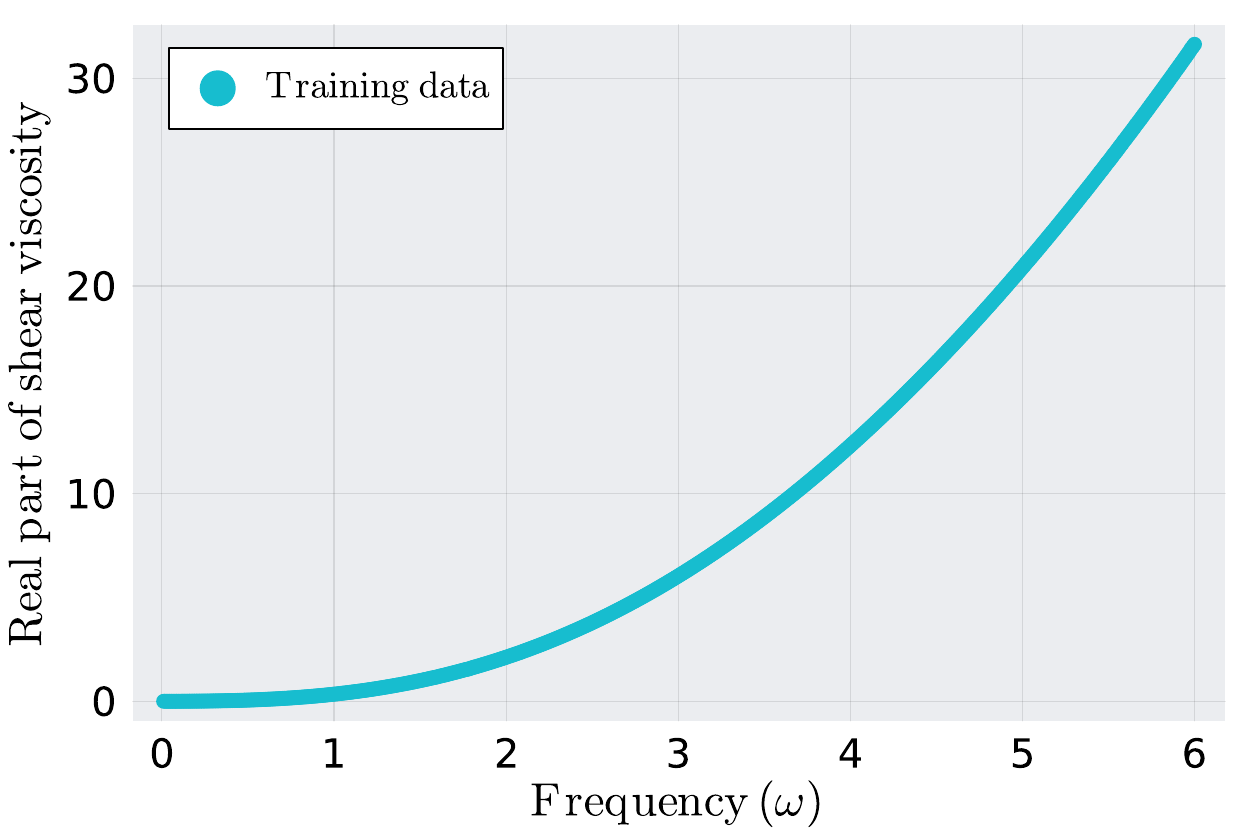}
  	\end{subfigure}
\begin{subfigure}{0.326\textwidth}
    	\includegraphics[width=\textwidth]{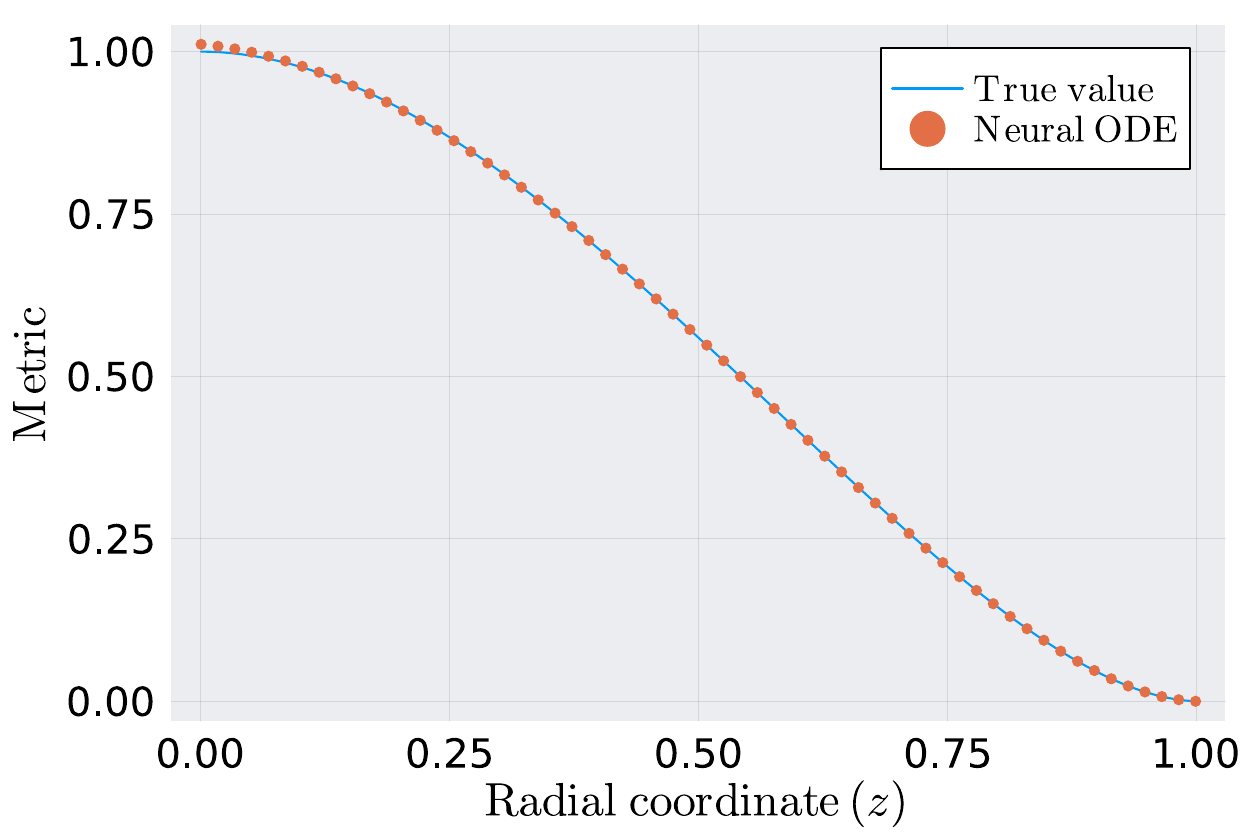}
  	\end{subfigure}
\begin{subfigure}{0.326\textwidth}
    	\includegraphics[width=\textwidth]{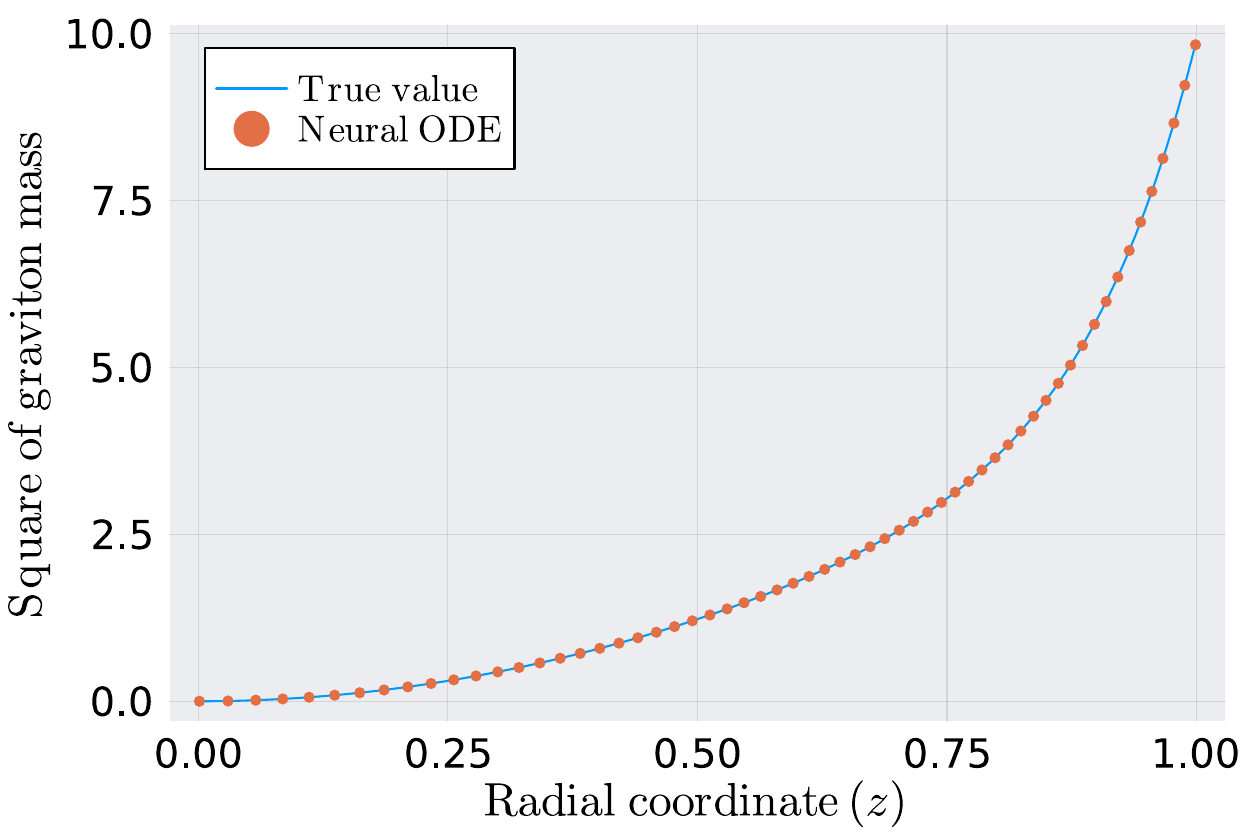}
  	\end{subfigure}
\caption{Training data and performance of Neural ODEs for the MG (top), LA (middle) and GA (bottom) model at
low temperatures.}
\label{fig:low_T}
\end{figure}

\begin{figure*}[h]  
\centering
\begin{subfigure}{0.326\textwidth}
    	\includegraphics[width=\textwidth]{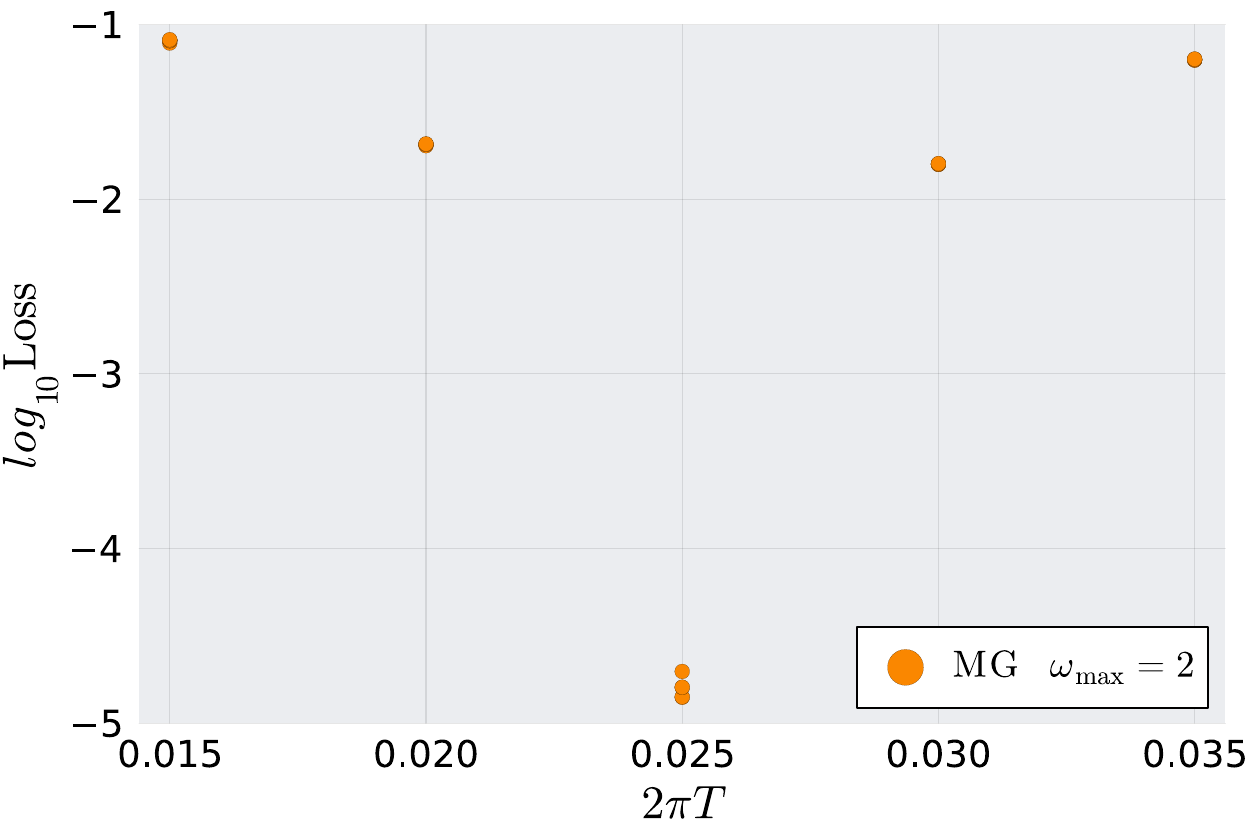}
  	\end{subfigure}
\begin{subfigure}{0.326\textwidth}
    	\includegraphics[width=\textwidth]{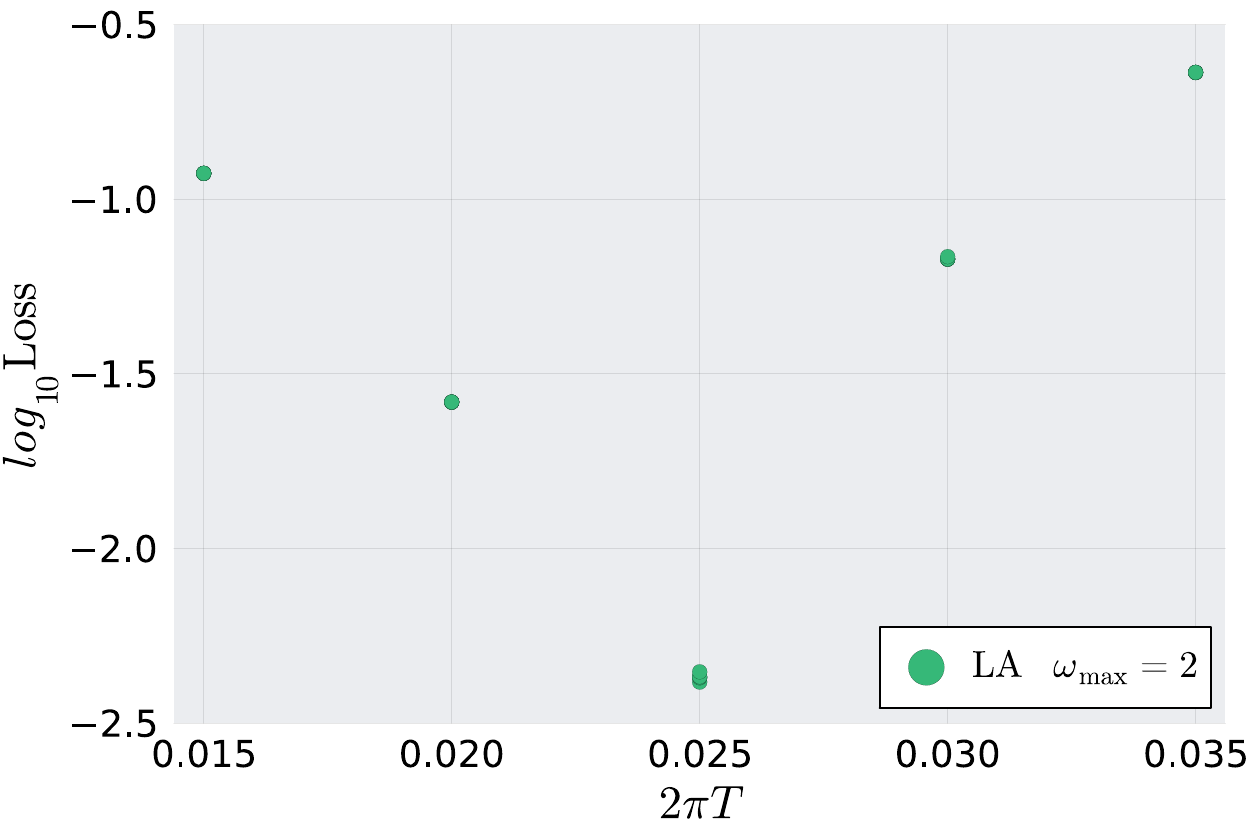}
  	\end{subfigure}
\begin{subfigure}{0.326\textwidth}
    	\includegraphics[width=\textwidth]{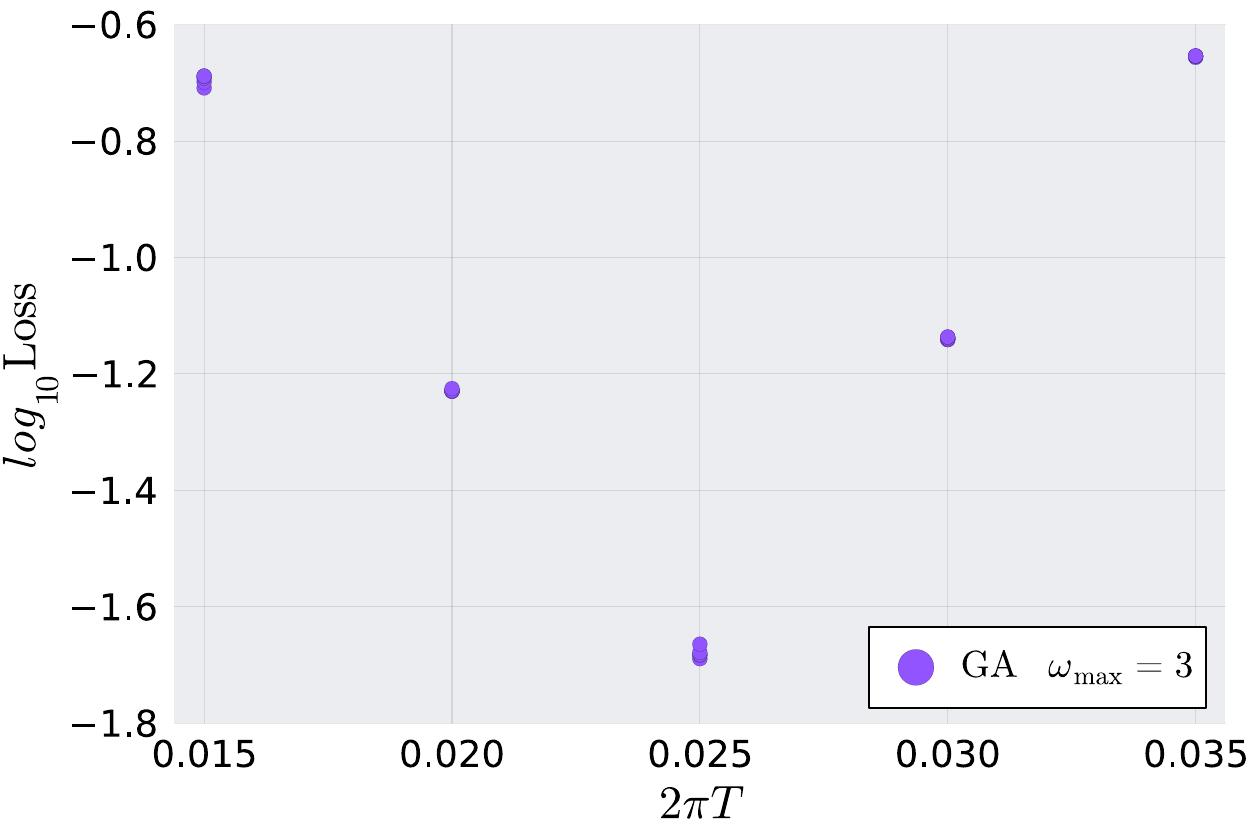}
  	\end{subfigure}   \\
\begin{subfigure}{0.326\textwidth}
    	\includegraphics[width=\textwidth]{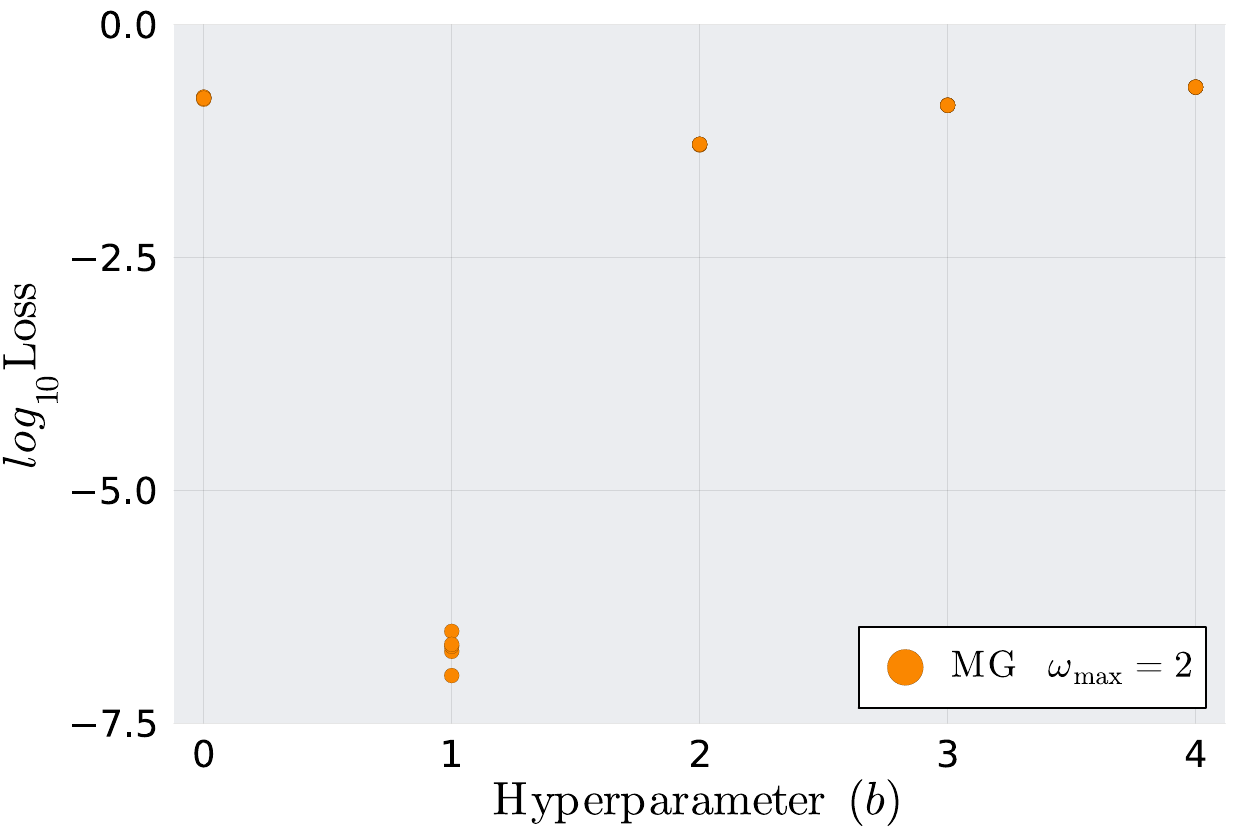}
  	\end{subfigure}
\begin{subfigure}{0.326\textwidth}
    	\includegraphics[width=\textwidth]{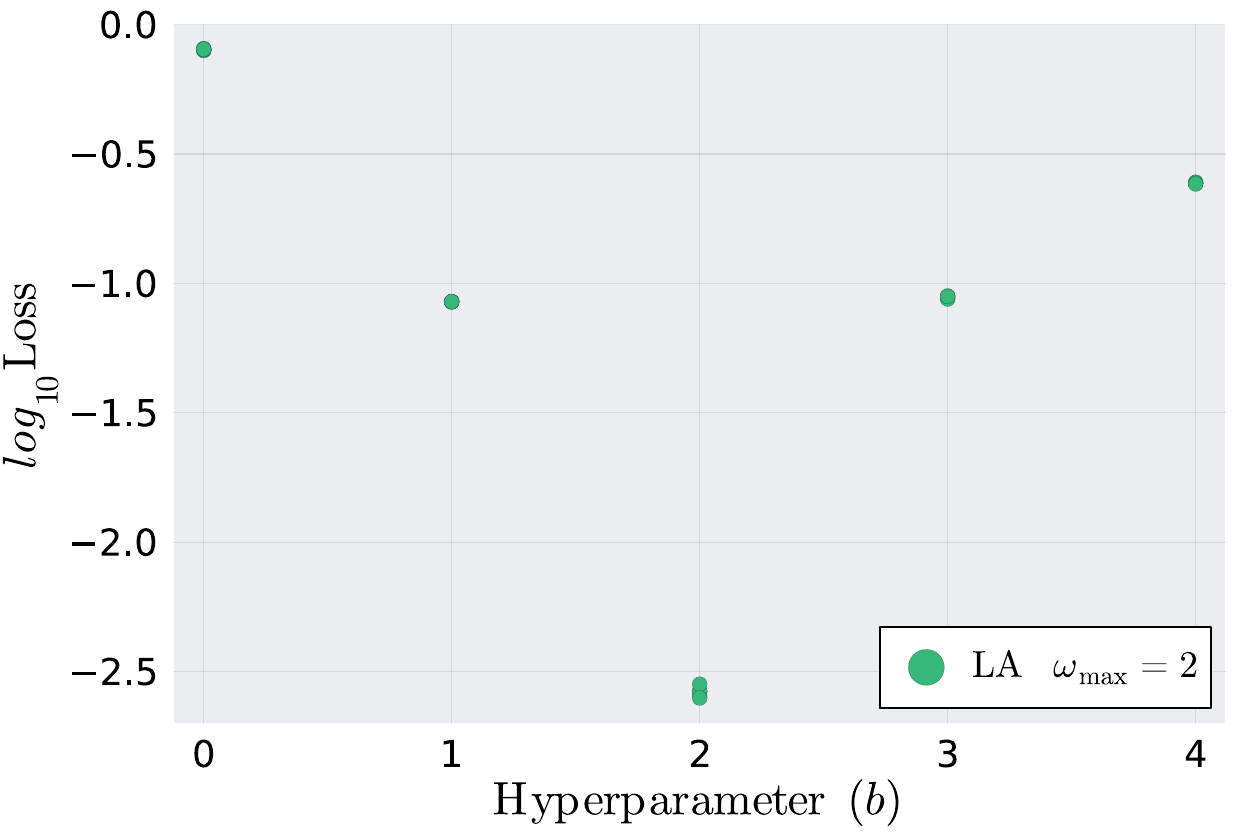}
  	\end{subfigure}
\begin{subfigure}{0.326\textwidth}
    	\includegraphics[width=\textwidth]{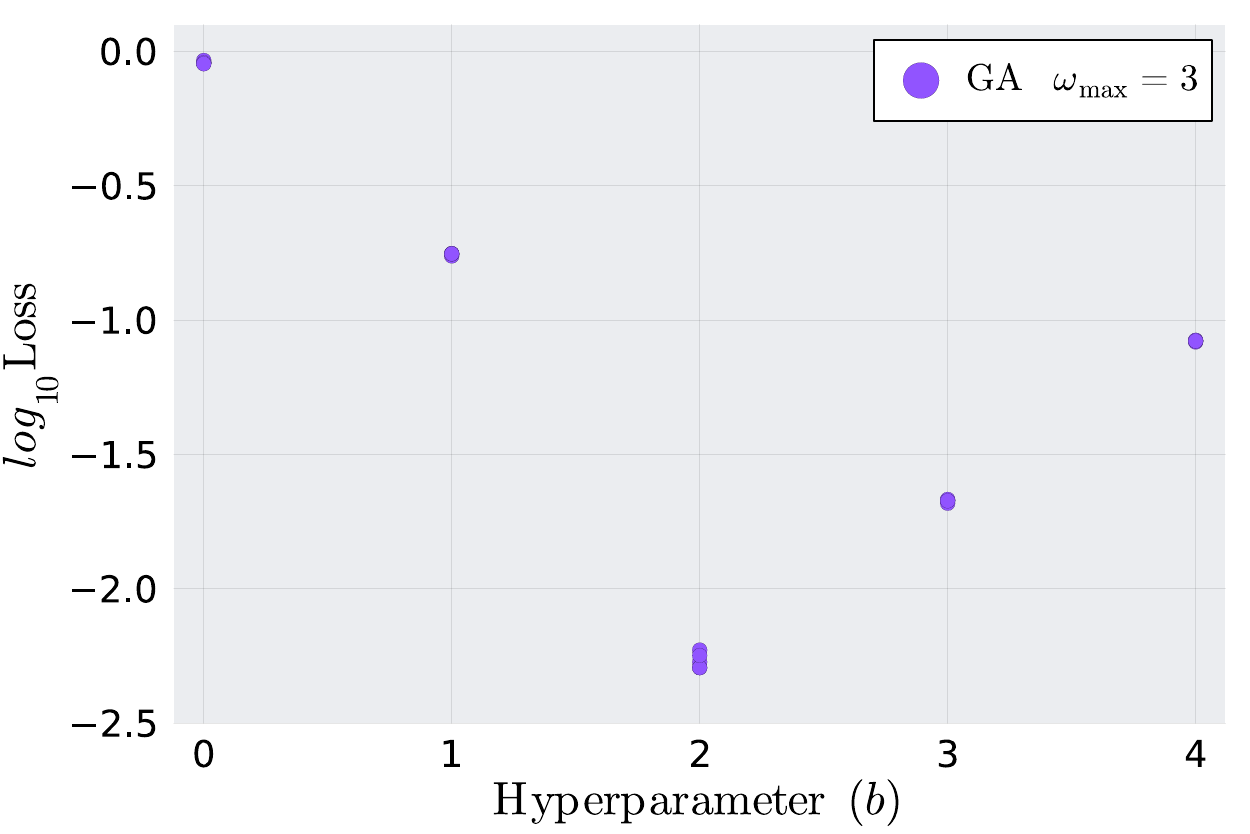}
  	\end{subfigure}
\caption{The hyperparameter turning of $2\protect\pi T$ (top) and $b$ (bottom) in three models at
low temperatures.}
\label{fig:Tb_lowT}
\end{figure*}

In order to learn the low-temperature physics, we will modify the algorithm
as follows.

1) Loss function

We will use a new loss function%
\begin{equation}
L=\frac{1}{N}\sum_{i=1}^{N}\frac{1}{2}\left[ \left\vert \frac{\eta _{\mathrm{%
re}}\left( \omega _{i},\theta \right) }{\bar{\eta}_{\mathrm{re}}\left(
\omega _{i}\right) }\right\vert -1-\log \left\vert \frac{\eta _{\mathrm{re}%
}\left( \omega _{i},\theta \right) }{\bar{\eta}_{\mathrm{re}}\left( \omega
_{i}\right) }\right\vert \right] ,
\end{equation}%
which is a combination of relative error and logarithmic difference. It has
a natural physical origin and will be elaborated in \cite{Ran2024}. Here we
introduce it just by experience.

2) Ansatz

We change the previous metric ansatz as%
\begin{equation}
f=(1-z)[4\pi T+(1-z)n_{1}(z,\theta _{1})],
\end{equation}%
where the temperature $T$ will be specified by hyperparameter tuning. Note
that this scheme reduces the difficulty of learning near-horizon geometry.

We also change the previous mass ansatz as%
\begin{equation}
m^{2}=z^{b}e^{n_{2}(z,\theta _{2})}.  \label{m22}
\end{equation}%
This is beneficial for learning the mass square in the GA model at low
temperatures, where $m^{2}/z^{2}$ grows fast near the horizon. Note that for
the low-temperature MG and LA models, as well as all the relatively
high-temperature models studied in the main text, the ansatz (\ref{m22})
remains applicable, and the performance of Neural ODEs will not be
significantly affected. The only price of using (\ref{m22}) is to assume
that the mass square is positive\footnote{%
Although there is no general argument, the mass square in all the models
given in \cite{Hartnoll1601} is positive, and the absence of instabilities
actually requires the positive mass square in MG and GA models \cite%
{Vegh1301,Baggioli2014}.}.

3) ODE solver

Since the Tsitouras 5/4 is less stable at low temperatures, we change the
solver to the Radau IIA5 \cite{Hairer1999}, which is a fully implicit
Runge-Kutta method and has high numerical stability.

4) Cutoffs

We set the IR and UV cutoffs as $z_{\mathrm{IR}}=0.999$ and $z_{\mathrm{UV}%
}=0.001$, by which the low-temperature data can still be generated with
sufficient accuracy.

5) Initial values

When learning the mass square of MG and LA models and tuning the
hyperparameters $T$ and $b$, we initialize the neural network with the
normal distribution $\mathcal{N}(0,0.001)$. For other tasks, we use $%
\mathcal{N}(0,1)$.

6) Hyperparameter tuning

The method of tuning $b$ remains the same as before. For tuning $T$, the
main difference is in the dataset separation: we randomly shuffle the
dataset within the frequency range $[0.01,\omega _{\max }]$ and split it
into two parts: 70\% for the training set and 30\% for the validation set.
Additionally, we remove the L1 regularization.

In figures \ref{fig:low_T}-\ref{fig:Tb_lowT} and table \ref{report2}, we
generate the data and illustrate the performance of Neural ODEs at low
temperatures. We choose the parameters ($2\alpha =5.9$, $\beta ^{2}=5.9$, $%
\gamma _{1}=4.9$, $\gamma _{5}=1$), which correspond to the same\
temperature $T\simeq 0.004$.

\begin{table}[h]
\centering%
\begin{tabular}{lllllll}
\hline\hline
Target & $f$ of MG & $m^{2}$ of MG & $f$ of LA & $m^{2}$ of LA & $f$ of GA & 
$m^{2}$ of GA \\ \hline
Loss & $3\times 10^{-7}$ & $<1\times 10^{-16}$ & $3\times 10^{-10}$ & $%
<1\times 10^{-16}$ & $9\times 10^{-8}$ & $2\times 10^{-8}$ \\ 
MRE & $1\times 10^{-3}$ & $8\times 10^{-10}$ & $2\times 10^{-4}$ & $2\times
10^{-9}$ & $1\times 10^{-3}$ & $1\times 10^{-4}$ \\ \hline\hline
\end{tabular}%
\caption{Minimum loss and MRE of six machine learning experiments at low
temperatures.}
\label{report2}
\end{table}

\section{Entanglement entropy}

\begin{figure*}[h]
\centering
\begin{subfigure}{0.326\textwidth}
    	\includegraphics[width=\textwidth]{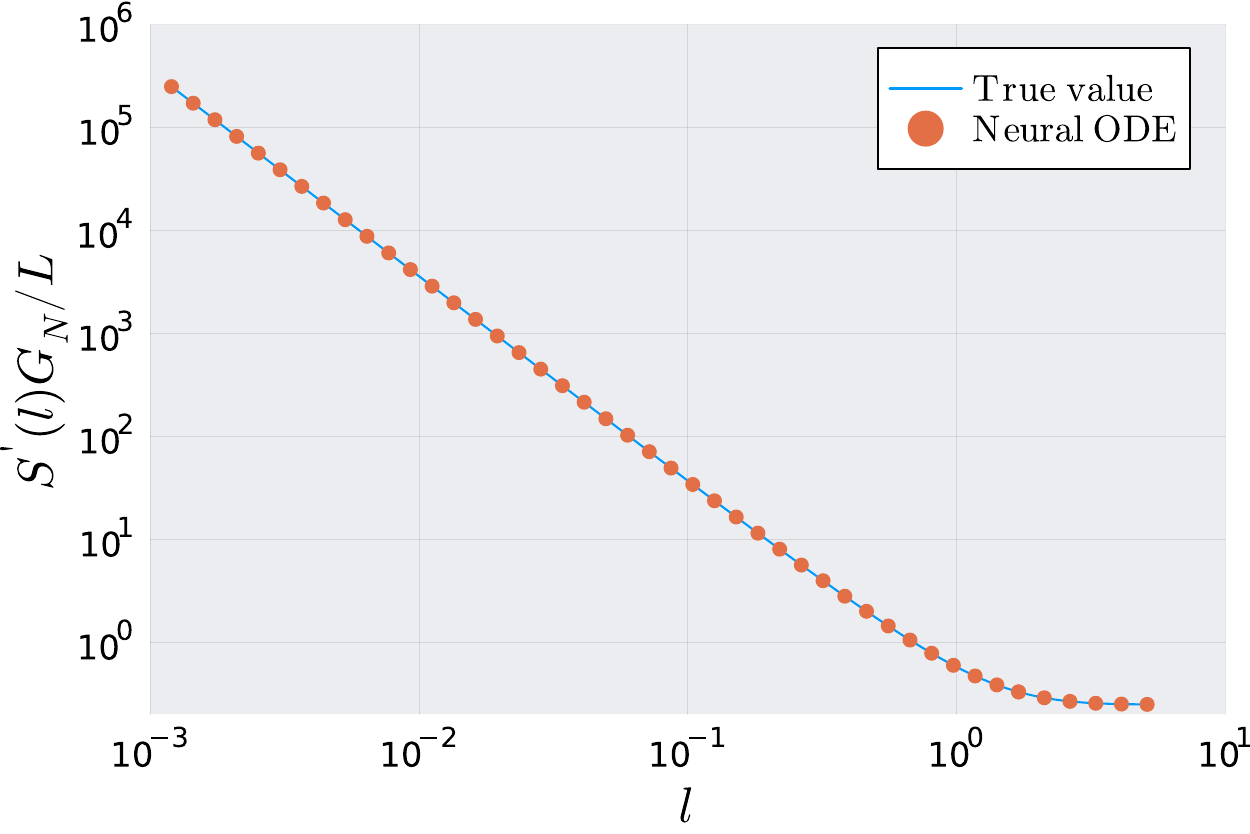}
  	\end{subfigure}
\begin{subfigure}{0.326\textwidth}
    	\includegraphics[width=\textwidth]{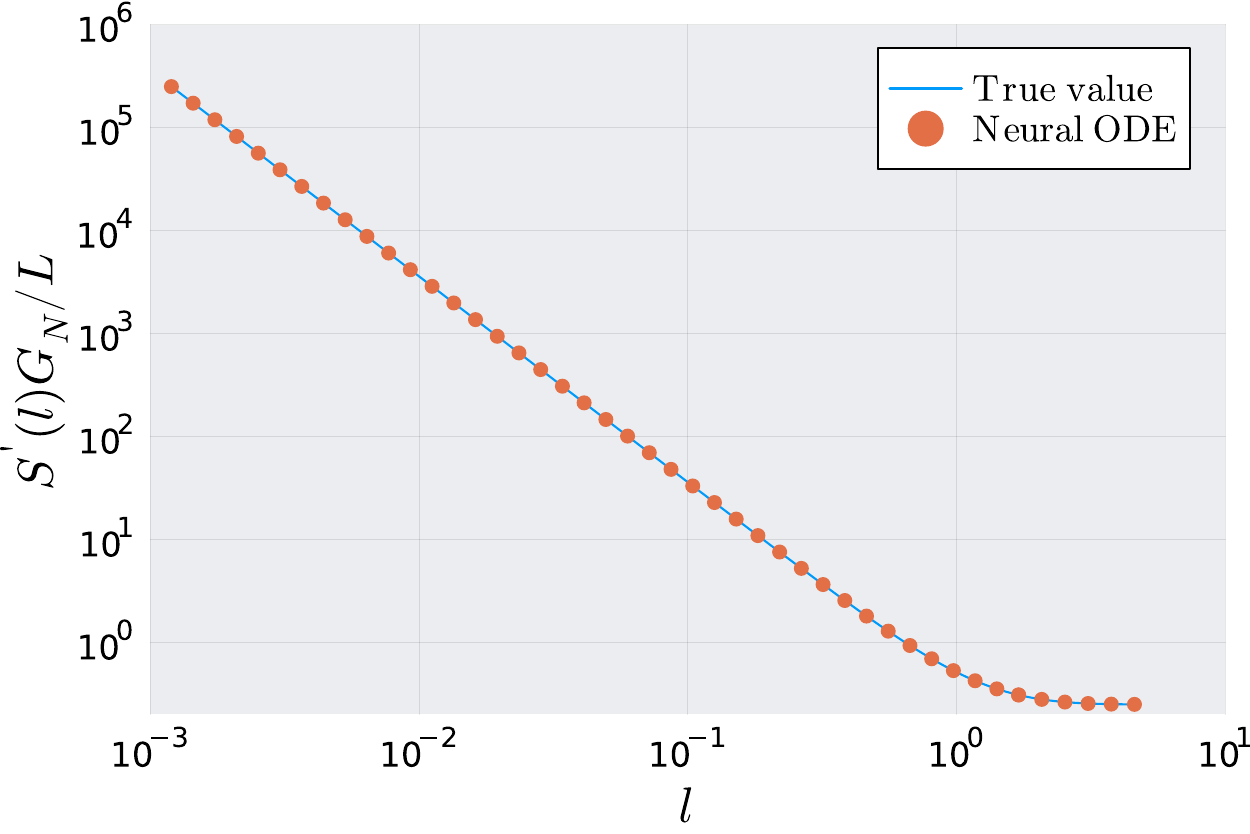}
  	\end{subfigure}
\begin{subfigure}{0.326\textwidth}
    	\includegraphics[width=\textwidth]{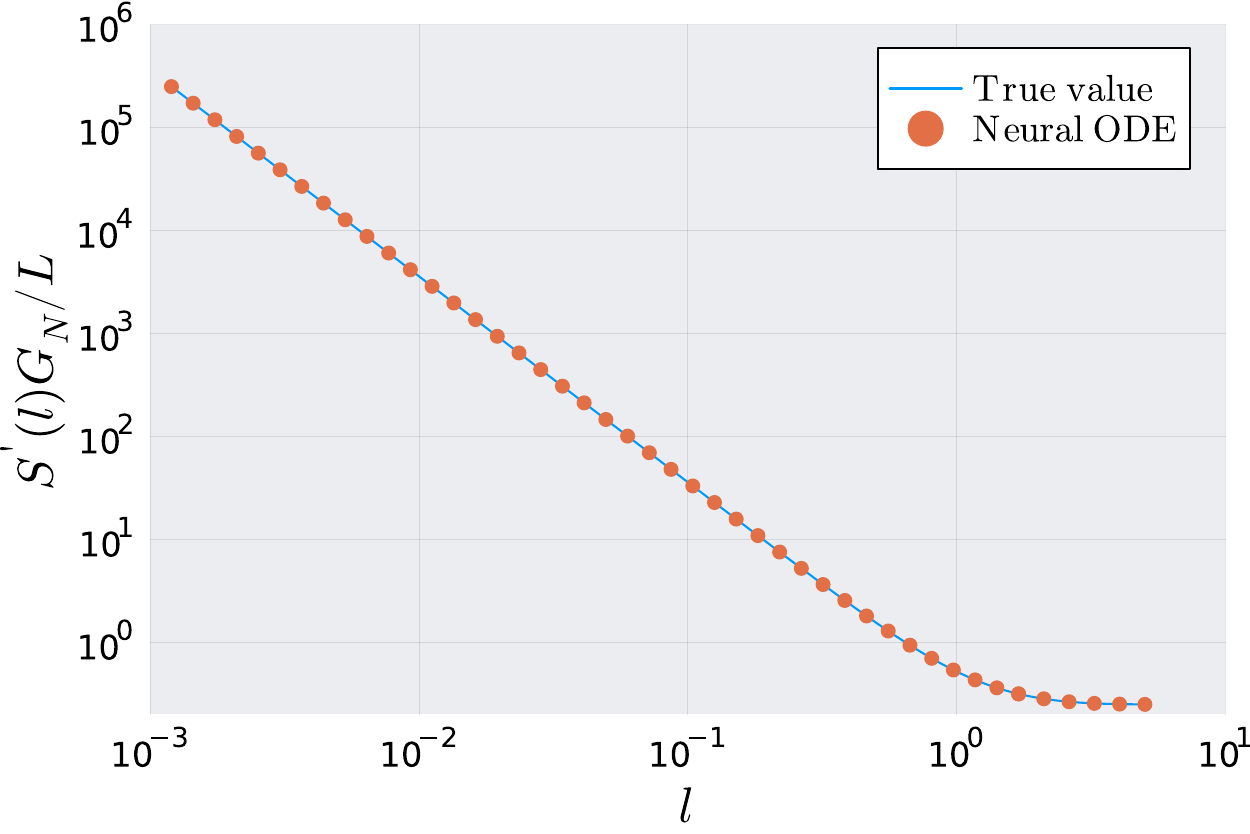}
  	\end{subfigure}  \\
\begin{subfigure}{0.326\textwidth}
    	\includegraphics[width=\textwidth]{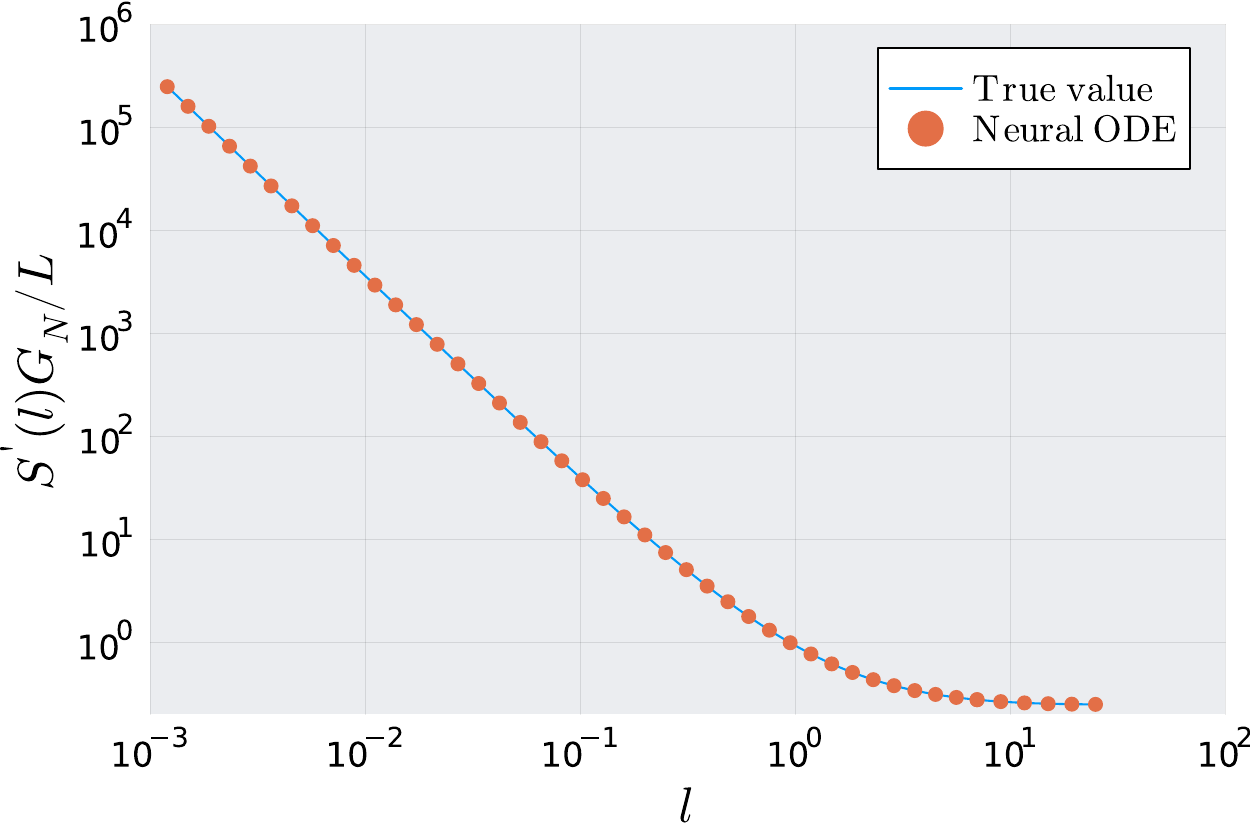}
  	\end{subfigure}
\begin{subfigure}{0.326\textwidth}
    	\includegraphics[width=\textwidth]{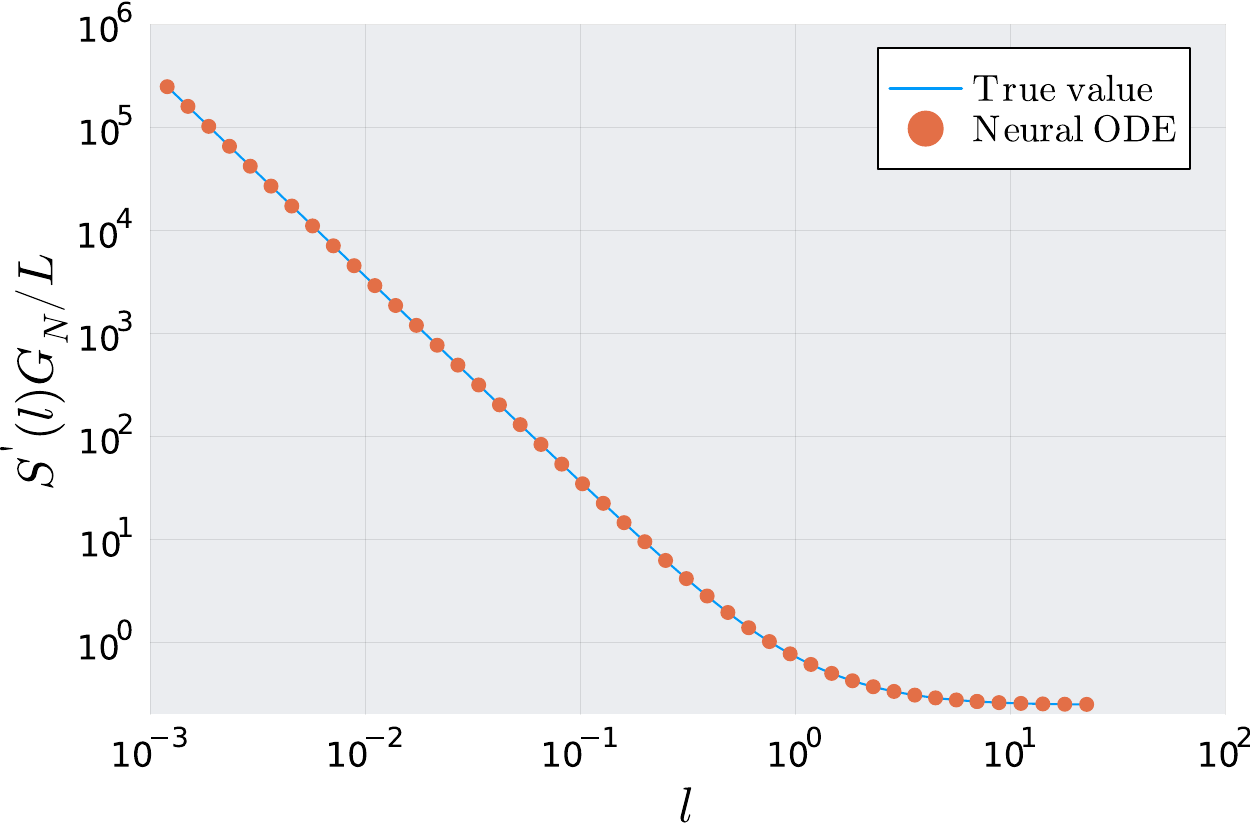}
  	\end{subfigure}
\begin{subfigure}{0.326\textwidth}
    	\includegraphics[width=\textwidth]{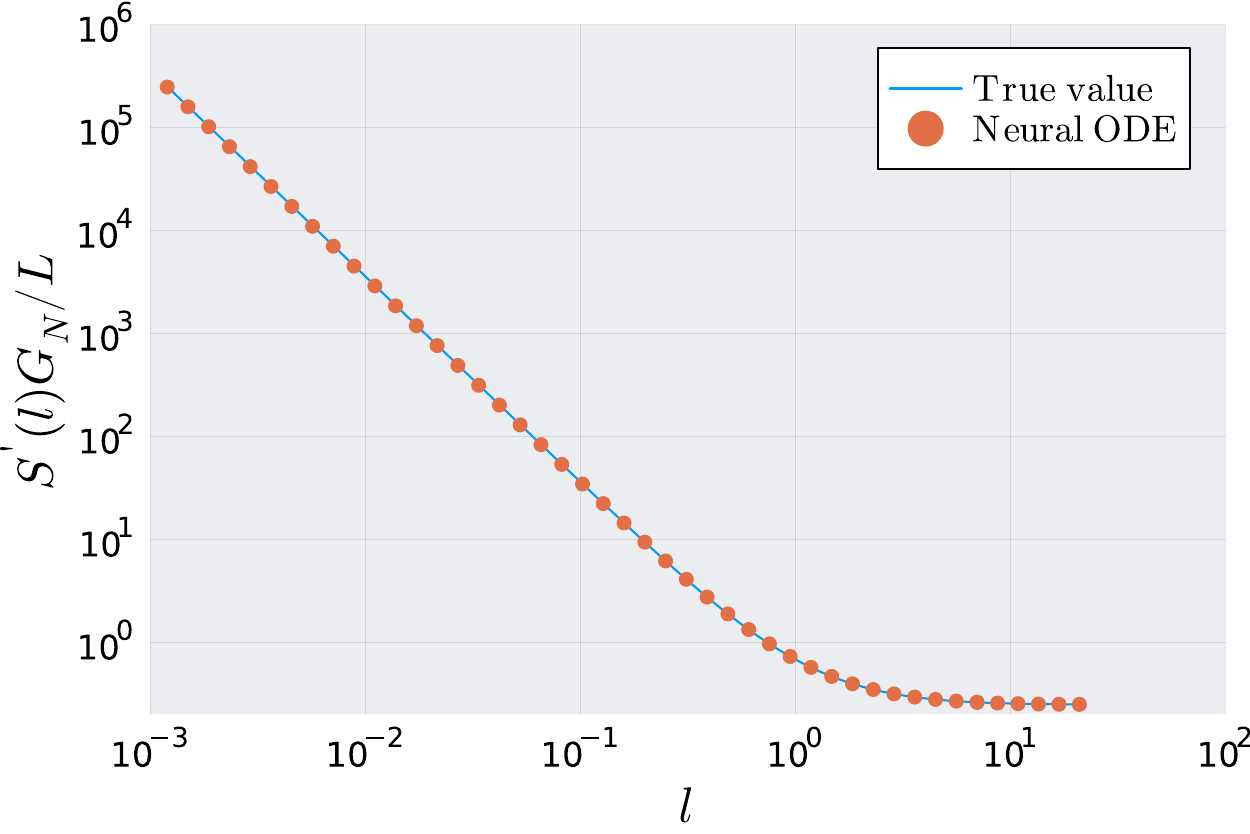}
  	\end{subfigure}
\caption{The performance of Neural ODEs in predicting $S^{\prime }(l)$. From
left to right, the three panels represent the results for the MG (left), LA (middle) and GA (right)
models at high (top) and low (bottom) temperatures.}
\label{fig:dS}
\end{figure*}

\begin{figure*}[h]
\centering
\begin{subfigure}{0.326\textwidth}
    	\includegraphics[width=\textwidth]{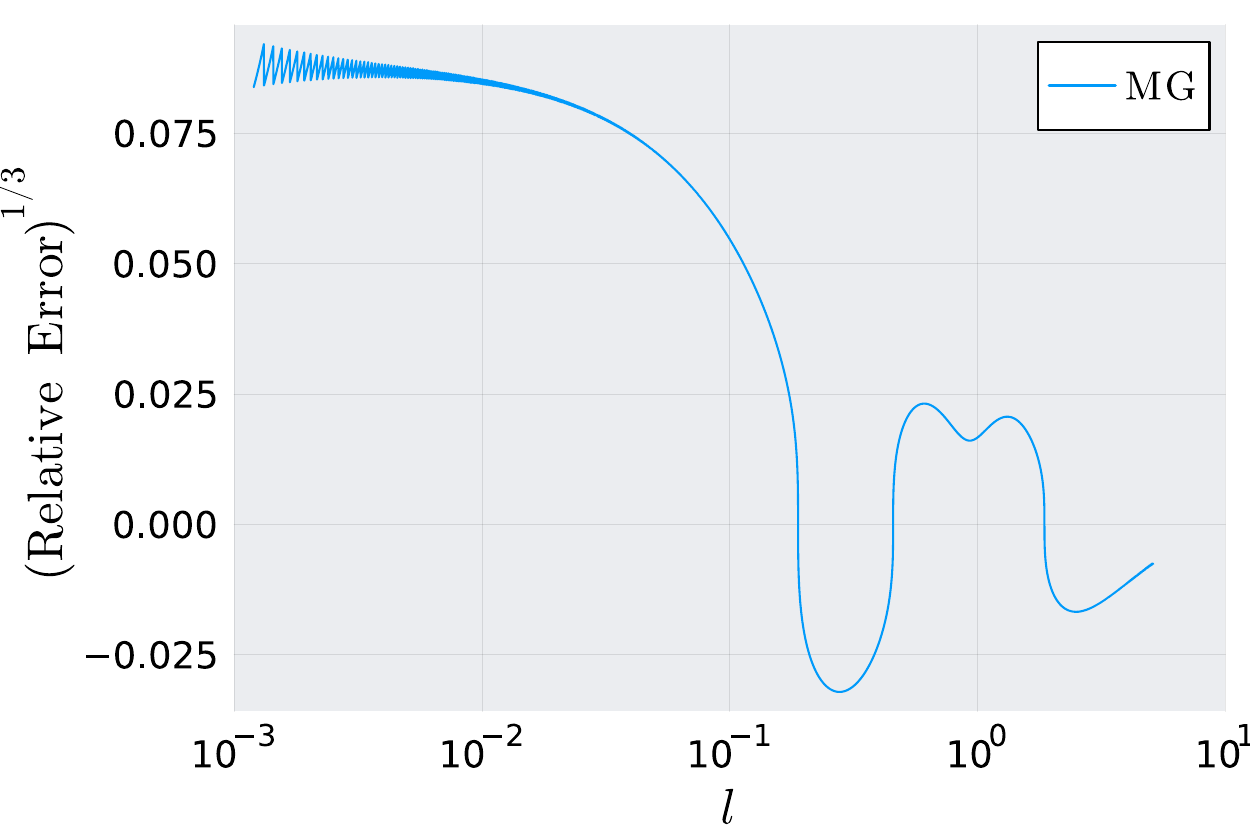}
  	\end{subfigure}
\begin{subfigure}{0.326\textwidth}
    	\includegraphics[width=\textwidth]{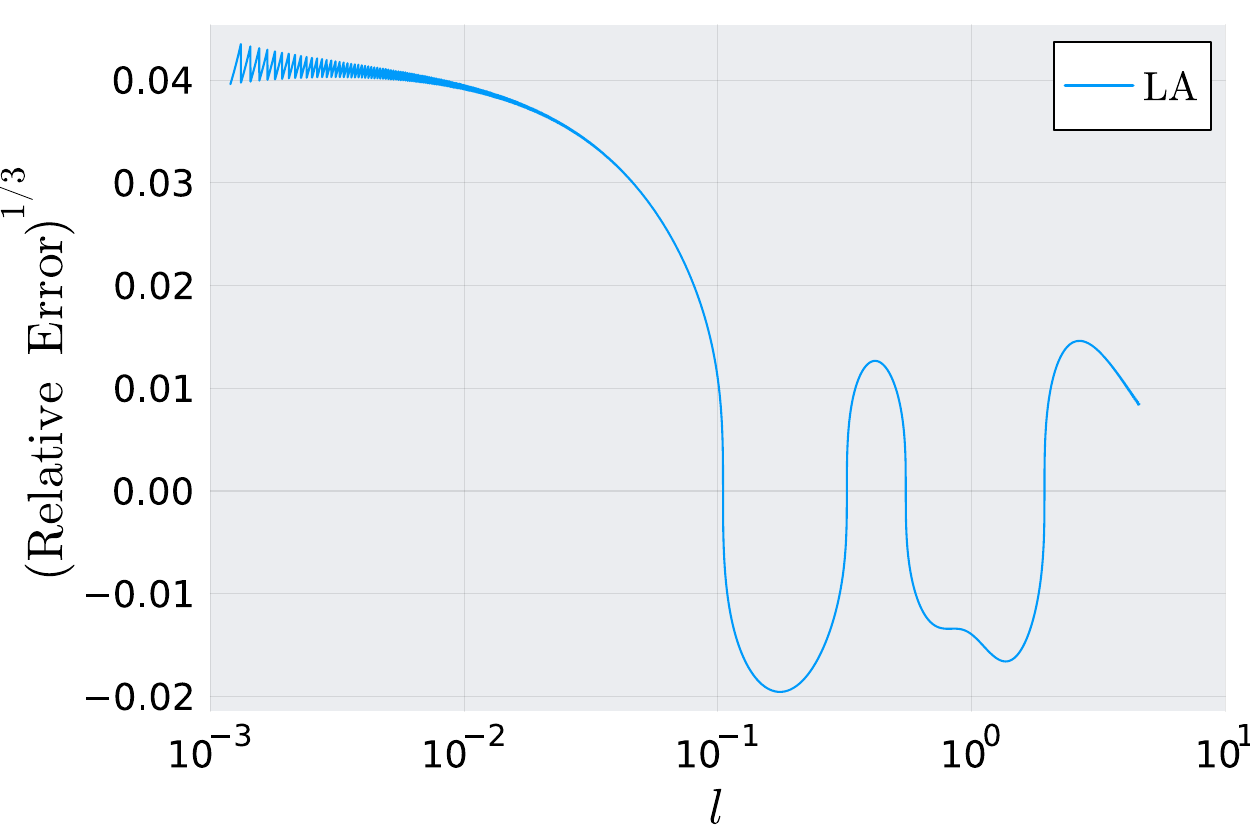}
  	\end{subfigure}
\begin{subfigure}{0.326\textwidth}
    	\includegraphics[width=\textwidth]{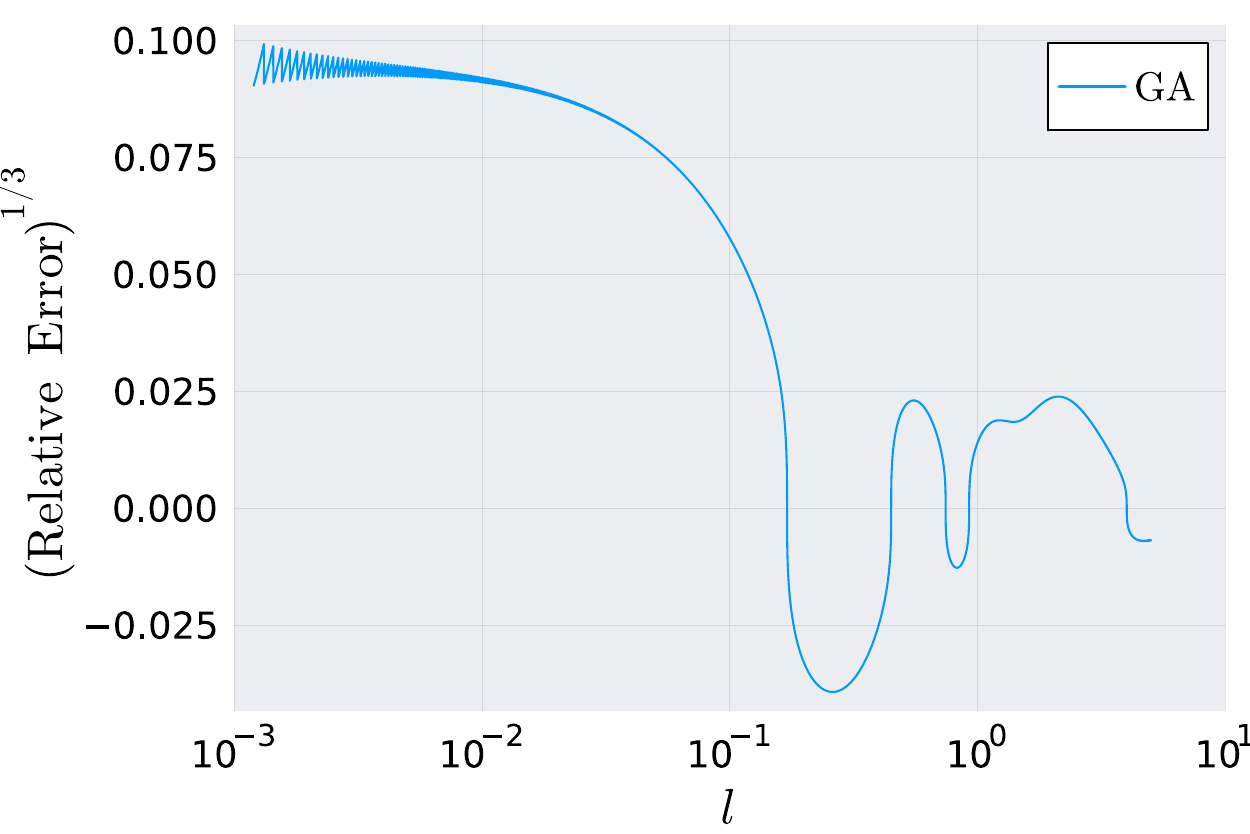}
  	\end{subfigure}   \\
\begin{subfigure}{0.326\textwidth}
    	\includegraphics[width=\textwidth]{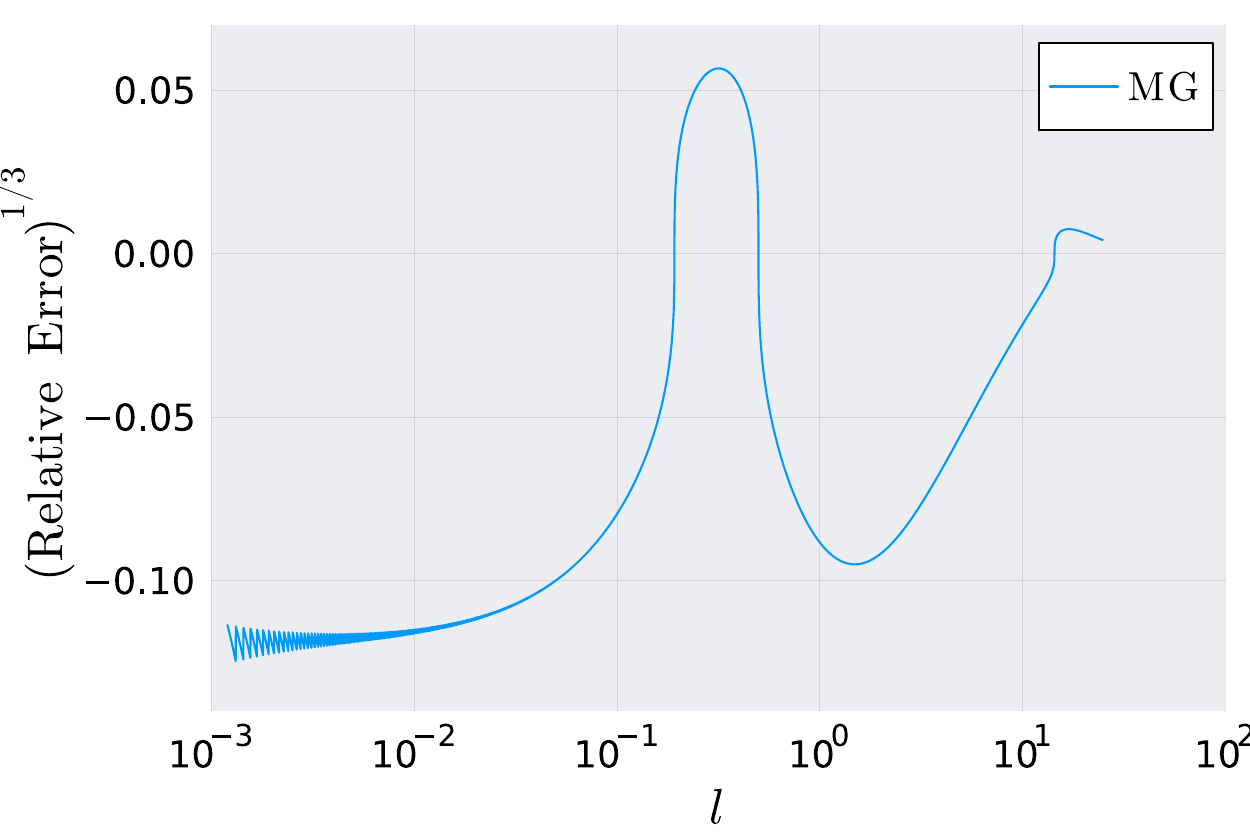}
  	\end{subfigure}
\begin{subfigure}{0.326\textwidth}
    	\includegraphics[width=\textwidth]{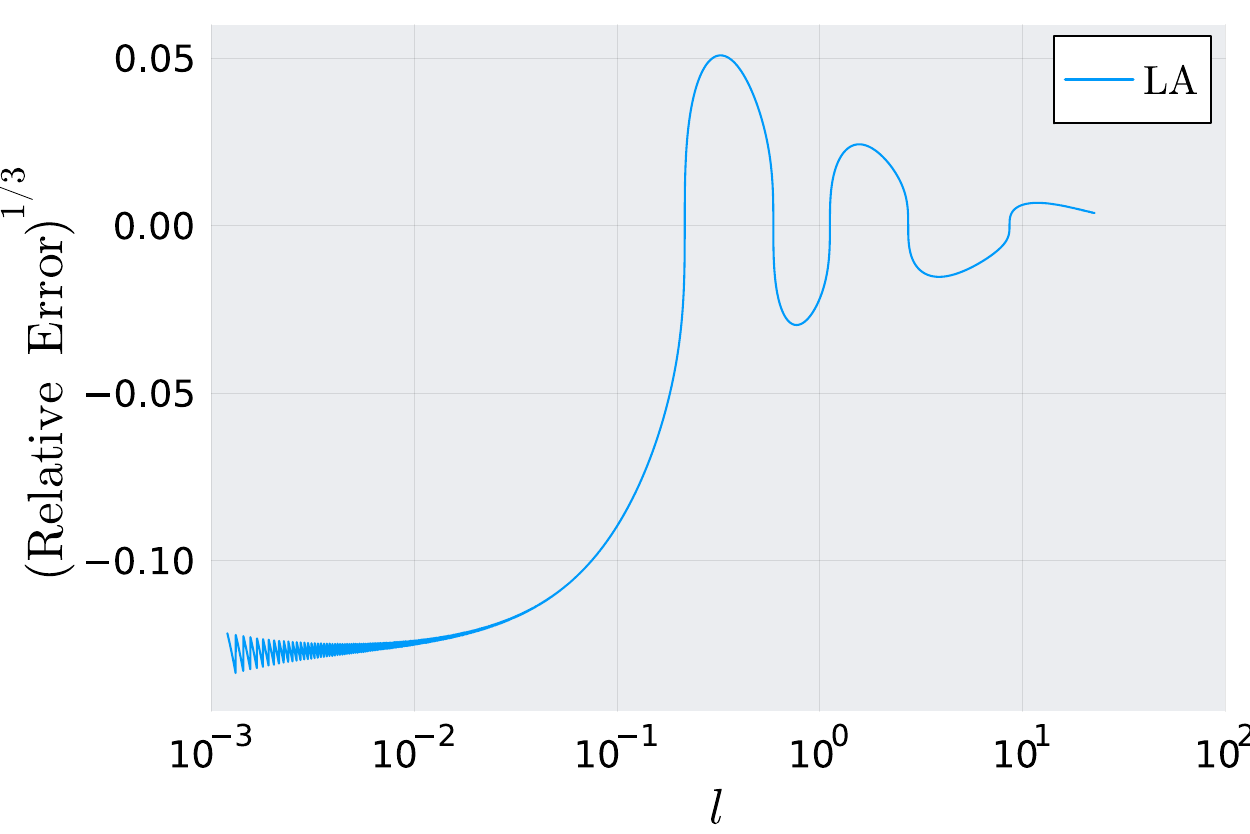}
  	\end{subfigure}
\begin{subfigure}{0.326\textwidth}
    	\includegraphics[width=\textwidth]{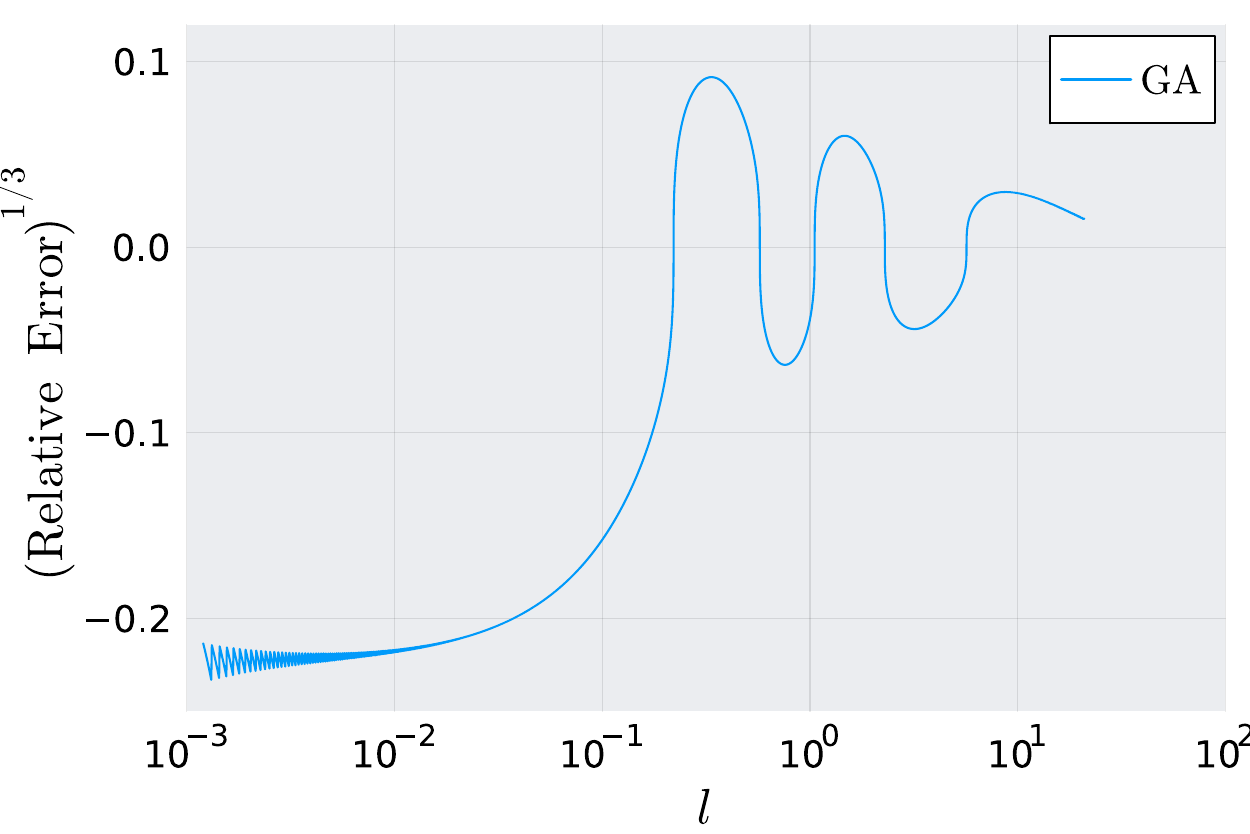}
  	\end{subfigure}
\caption{Relative errors in $S^{\prime }(l)$ for three models at high (top) and low (bottom) temperatures.}
\label{fig:err}
\end{figure*}

At last, we will use the bulk metric learned by Neural ODEs to predict the
derivative of EE with respect to the size of entangling region.

Consider the region $A$ to be a spatial strip with infinite length $L$ and
width $l$ on the boundary. The EE of $A$ can be calculated by the RT formula 
\cite{Ryu:2006bv}%
\begin{equation}
S=\frac{\mathrm{Area}(\gamma _{A})}{4G_{N}},  \label{eq:sec 2 SS}
\end{equation}%
where $\gamma _{A}$ is the minimal surface in the bulk that is homologous to
the boundary region $A$. Using the profile equation for $\gamma _{A}$, the
EE and the width can be expressed as \cite{Bilson:2010ff}%
\begin{eqnarray}
S(z_{\ast }) &=&\frac{L}{2G_{N}}\int_{\delta }^{z_{\ast }}\frac{z_{\ast }^{2}%
}{z^{2}\sqrt{f(z)}\sqrt{z_{\ast }^{4}-z^{4}}}dz,  \label{eq:sec 2 Sf} \\
\,l(z_{\ast }) &=&2\int_{0}^{z_{\ast }}\frac{z^{2}}{\sqrt{f(z)}\sqrt{z_{\ast
}^{4}-z^{4}}}dz,  \label{eq:sec 2 lf}
\end{eqnarray}%
where $z_{\ast }$ is the radial coordinate at the tip of $\gamma _{A}$, and $%
\delta $ is the UV cutoff that is introduced to avoid the divergence of EE.
Comparing the derivatives of above two equations%
\begin{eqnarray}
\frac{dS(z_{\ast })}{dz_{\ast }} = \frac{L}{2G_{N}}\Bigg[ \frac{1}{\sqrt{%
f(z_{\ast })}}\lim_{z\rightarrow z_{\ast }}\frac{1}{\sqrt{z_{\ast }^{4}-z^{4}%
}} -\int_{0}^{z_{\ast }}\frac{2z^{2}z_{\ast }}{\left( z_{\ast
}^{4}-z^{4}\right) ^{\frac{3}{2}}\sqrt{f(z)}}dz\Bigg] , \\
\frac{dl(z_{\ast })}{dz_{\ast }} = 2z_{\ast }^{2}\Bigg[ \frac{1}{\sqrt{%
f(z_{\ast })}}\lim_{z\rightarrow z_{\ast }}\frac{1}{\sqrt{z_{\ast }^{4}-z^{4}%
}}-\int_{0}^{z_{\ast }}\frac{2z^{2}z_{\ast }}{\left( z_{\ast
}^{4}-z^{4}\right) ^{\frac{3}{2}}\sqrt{f(z)}}dz\Bigg] ,
\end{eqnarray}%
where we have set $\delta\rightarrow 0$, one can obtain a simple formula%
\begin{equation}
\frac{dS(l)}{dl}=\frac{L}{4G_{N}z_{\ast }^{2}}.  \label{eq:sec 2 dSdl}
\end{equation}

With these preparations in place, let's substitute the bulk metric (the
learned metric or the true value) into eq. (\ref{eq:sec 2 lf}) and get $%
l(z_{\ast })$ by numerical integration. By the interpolation, we can
construct the inverse function $z_{\ast }(l)$. In terms of eq. (\ref{eq:sec
2 dSdl}) with $z_{\ast }(l)$, the derivative $S^{\prime }(l)$ can be
obtained. In figure (\ref{fig:dS}), we
plot $S^{\prime }(l)$ for the three models at (relatively) high temperatures
and low temperatures. The change of the relative error along $l$ is depicted
in figure (\ref{fig:err}). Notably, the relative error decreases oscillatingly as $l$ increases. At last, we calculate the MRE by uniformly sampling 10000 points within the allowed
range of $l$, which are given in Table \ref{report3}. 

\begin{table}[h]
\centering%
\begin{tabular}{llll}
\hline\hline
Model & MG & LA & GA \\ \hline
High temperature & $1\times 10^{-5}$ & $3\times 10^{-6}$ & $2\times 10^{-5}$
\\ 
Low temperature & $1\times 10^{-4}$ & $1\times 10^{-5}$ & $8\times 10^{-5}$
\\ \hline\hline
\end{tabular}%
\caption{The MRE of $S^{\prime }(l)$ for three models at two temperatures.}
\label{report3}
\end{table}

\end{document}